\documentclass[fleqn,10pt]{wlscirep}
\usepackage[utf8]{inputenc}
\usepackage[T1]{fontenc}
\usepackage{mathtools}
\usepackage{times}
\usepackage{lineno}
\usepackage{xcolor}
\usepackage{xspace}
\usepackage{hyperref}
\usepackage{comment}

\usepackage{pdfpages}

\newcommand{\ie}{\emph{i.e.}}
\newcommand{\eg}{\emph{e.g.}}

\newcommand{\ER}{Erd\"{o}s-R\'enyi }
\newcommand{\avg}[1]{\langle #1\rangle}
\newcommand{\mx}[1]{\mathbf{#1}}
\newcommand{\mxt}[1]{\tilde{\mathbf{#1}}}

\title{Meta-validation of bipartite network projections}

\author[1,2,*]{Giulio Cimini}
\author[3]{Alessandro Carra}
\author[3]{Luca Didomenicantonio}
\author[4,2]{Andrea Zaccaria}
\affil[1]{Physics Department and INFN, University of Rome Tor Vergata, 00133 Rome (Italy)}
\affil[2]{Enrico Fermi Research Center, 00184 Rome (Italy)}
\affil[3]{Physics Department, Sapienza University of Rome, 00185 Rome (Italy)}
\affil[4]{Institute for Complex Systems (CNR) UoS Sapienza, 00185 Rome (Italy)}
\affil[*]{giulio.cimini@roma2.infn.it}

\date{}

\begin{abstract}
Monopartite projections of bipartite networks are useful tools for modeling indirect interactions in complex systems. 
The standard approach to identify significant links is statistical validation using a suitable null network model, such as the popular configuration model (CM) that constrains node degrees and randomizes everything else. However different CM formulations exist, depending on how the constraints are imposed and for which sets of nodes.
Here we systematically investigate the application of these formulations in validating the same network, showing that they lead to different results even when the same significance threshold is used. Instead a much better agreement is obtained for the same density of validated links.
We thus propose a meta-validation approach that allows to identify model-specific significance thresholds for which the signal is strongest, and at the same time to obtain results independent of the way in which the null hypothesis is formulated. We illustrate this procedure using data on scientific production of world countries.
\end{abstract}

\begin{document}

\flushbottom
\maketitle
\thispagestyle{empty}

\section*{Introduction}

Networks are simplified yet effective models for a large class of natural, socio-economic and technological systems described by complex interaction patterns. 
Independently of the nature of the underlying interactions, the network representation allows capturing the emergent features of these systems as well as their dynamical patterns \cite{song2005self,boccaletti2006complex,dorogovtsev2008critical,pastorsatorras2015epidemic,benson2016higher}.
As such, network science has gained increasing popularity in the last twenty years \cite{barabasi2012network,newman2018networks,caldarelli2020perspective}. 

A network is labeled as \emph{bipartite} when its elements (the nodes) can be split in two disjoint sets, such that links can only exist between nodes of different sets \cite{holme2003network}. Bipartite networks are the natural representation for several systems, such as: 
social  affiliation and collaboration networks, where individuals connect to the groups they are member of \cite{faust1997centrality,newman2004coauthorship}; financial and commercial ownership networks, where entities are linked to the goods they own or consume \cite{zhou2007bipartite,bardoscia2021physics}; trade networks, where economies connect to the products they export \cite{hidalgo2009building,tacchella2012new}; ecological networks, where species connect to the habitat they live in \cite{ings2009review,mariani2019nestedness}; biological and medical networks connecting, \eg, patients and diseases \cite{goh2007human,pavlopoulos2018bipartite}.
Mathematically speaking, a bipartite network is defined as a graph with two sets $\text{L}$ and $\Gamma$ of nodes, and a $|\text{L}|\times|\Gamma|$ matrix of connections $\mx{M}$ called \emph{bi-adjacency matrix}. 
The generic element of this matrix is 
\begin{equation}
M_{i\alpha}=
\begin{cases}
1\mbox{ if nodes $i\in\text{L}$ and $\alpha\in\Gamma$ are connected,}\\
0\mbox{ otherwise.}
\end{cases}
\label{eq:def_bipa}
\end{equation}
The number of connections or \emph{degree} of a node $i\in\text{L}$ is then defined as $k_i = \sum_{\alpha\in\Gamma} M_{i\alpha}$, while the degree of a node $\alpha\in\Gamma$ is $\kappa_\alpha=\sum_{i\in\text{L}} M_{i\alpha}$. The total number of links in the network is $E=\sum_{i\in\text{L}}k_i=\sum_{\alpha\in\Gamma}\kappa_\alpha$.

The indirect relation between two nodes belonging to the same set of a bipartite network can be measured through their \emph{co-occurrences} (or common neighbors), namely how many nodes of the other set they are both connected to. 
For instance the co-occurrences of nodes $i$ and $j$ of set $\text{L}$ are given by
\begin{equation}
C_{ij}=\sum_{\alpha\in\Gamma} M_{i\alpha} M_{j\alpha}.
\label{eq:def_proj}
\end{equation}
The $\text{L}\times\text{L}$ square matrix $\mx{C}$ represents a monopartite network obtained as the projection of the original bipartite network onto the set $\text{L}$ ($\text{L}$-projection) \cite{zhou2007bipartite}. 
Analogously, one can project the bipartite network onto the set $\Gamma$ to obtain the co-occurrences between nodes of that set ($\Gamma$-projection). 

The main problem in studying bipartite network projections is that they are often very dense and thus difficult to handle with the tools of network theory. This happens because any two nodes are connected in the projected network as soon as they have a single co-occurrence. Moreover, co-occurrences can be influenced by single node variables, thus understanding whether they indicate an effective interdependence between nodes may be difficult. For example, nodes that have high degree in the bipartite network naturally tend to have more co-occurrences than low-degree nodes (more generally, the degree sequence of the network projection is highly dependent on the degree sequence of the two sets from the original bipartite structure \cite{vasquez2018degree}). 
It is thus useful to extract representative links of the projected network; this can be achieved using several filtering techniques, from unconditional thresholding to Minimal Spanning Trees \cite{kruskal1956shortest} and Planar Maximally Filtered Graphs \cite{tumminello2005tool}.
Yet in order to identify the most informative co-occurrences, the statistically-grounded approach consists in performing link validation using a null network model. 

Statistical validation of network patterns is a common approach in the literature (the classical applications being  motifs expression analysis \cite{maslov2002specificity,milo2002network}, network backbone extraction \cite{serrano2009extracting} and community detection \cite{fortunato2016community,macmahon2015community,bongiorno2017core}). The goal is to identify the empirical patterns that deviate from a benchmark null model, in order to ensure that those patterns are indeed a salient feature of the network and not a mere consequence of some of its other properties (given the potentially strong interdependence between structural network quantities \cite{vazquez2004topological,foster2011clustering,colomer2013deciphering,orsini2015quantifying,marcaccioli2019polya}). 
Following the prescription of information theory \cite{cimini2019statistical}, the null model is thus obtained by constraining some network properties and randomizing everything else. In this way, the formulated null hypothesis is that these constraints are the only explanatory variables for the network at hand; when the null hypothesis is rejected, we can state that the observed network patterns are not a mere consequence of the imposed constraints. 

Going back to our context of bipartite network projections, the statistical significance of each observed co-occurrence value $C_{ij}>0$ can be quantified through its p-value:
\begin{equation}
p[C_{ij}] = 1-\sum_{x=0}^{C_{ij}-1}\pi(x|i,j),
\label{eq.pval}
\end{equation}
where $\pi(\cdot|i,j)$ is the probability distribution of the expected co-occurrences between $i$ and $j$ under the null model.
The right-hand side of eq.~\eqref{eq.pval} is the probability that $i$ and $j$ have no less than $C_{ij}$ co-occurrences in the null model. This quantity can be used to build a \emph{validated} (or filtered) projection of the original bipartite network, containing only the most significant links according to the null model. For each $C_{ij}$, if the p-value of eq.~\eqref{eq.pval} is smaller than a significance threshold (or confidence level) $p^*$, the link $i,j$ is placed on the monopartite validated network; otherwise, it is discarded. 
In other words, the comparison is deemed statistically significant if the observed co-occurrences are an unlikely realization of the null hypothesis according to the significance level $p^*$ (in particular, we are interested in detecting the co-occurrences that are significantly larger than their null model expectation; significantly smaller values can be obtained in a similar fashion --- see Supplementary Note 1).  
In this way the original amount of links is drastically reduced, and the result is a much sparser validated network with a clearer meaning. 

Naturally, statistical validation has some intrinsic degrees of freedom: the choice of the null model, its specific formulation, and the value of the significance level $p^*$. In particular, the choice of the model is a step that should be handled with care, as a bad choice may lead to wrong conclusions about the structural and functional features of the network \cite{colizza2006detecting,amaral2006lies}.
For instance, using a (bipartite) \ER model \cite{erdos1959random}, \ie, a random network preserving only the density of the original bipartite graph, leads to an identical distribution $\pi(\cdot|i,j)$ for each node pair $i,j$ and thus to an unconditional global threshold to select the most significant $C_{ij}$ values \cite{latapy2008basic,neal2013identifying}. However this choice does not solve the bias problem for high degree nodes, which is very important in networks given that degree distributions are typically very broad \cite{serafinoe2021true}. 
A natural way to take this aspect into account is given by the popular \emph{configuration model} (CM)\cite{chung2002connected,newman2001random,squartini2011analytical}, which generates random networks with a given degree sequence. 

In the context of bipartite networks, the first model formulation of this family was obtained by constraining the degrees of nodes in one set (say $\text{L}$) \cite{teece1994understanding}. In this case, the co-occurrences probability can be computed exactly as a hypergeometric distribution \cite{goldberg2003assessing,tumminello2011statistically}. Yet this model solves the degree bias only partially, since it assumes nodes of the other set (say $\Gamma$) to be equivalent and interchangeable. 
The alternative approach is to model random bipartite networks preserving the degrees of both node sets $\text{L}$ and $\Gamma$, and then use these networks to obtain the null model for the projected network. 
\emph{Degree sequence models} follow this approach, however they either require multiple observations of the empirical network \cite{neal2014backbone} or are based on computational link swap methods \cite{zweig2011systematic} that are typically impractical and biased \cite{gionis2007assessing}. 
An exception is represented by the recently proposed \emph{Curveball} algorithm \cite{verhelst2008efficient,strona2014fast,carstens2015proof}, a link swap method that is extremely efficient in generating network configurations and is ergodic (\ie, it can sample uniformly over the set of all possible network configurations). 
The alternative route to Monte Carlo sampling is represented by maximum-entropy models. The \emph{Bipartite Configuration Model} (BiCM) \cite{saracco2015randomizing} allows generating an ensemble of bipartite networks where node degrees of both sets $\text{L}$ and $\Gamma$ are preserved as  ensemble expectations. The null model for the network projections is then obtained by projecting BiCM-generated networks \cite{gualdi2016statistically,saracco2017inferring}. 
This latter approach allows computing the co-occurrences distributions both numerically and analytically, and simplifies as a \emph{Bipartite Partial Configuration Model} (BiPCM) when degree constraints are imposed only on one set of nodes.
At last we note that, in principle, the projection of a bipartite network can be statistically validated also using a null model for monopartite weighted networks \cite{serrano2009extracting,mastrandrea2014enhanced,marcaccioli2019polya}. 
That is, instead of defining the null model on the bipartite network and then deriving its formulation for the network projection, the null model can be directly defined on the monopartite projection. However this approach discards the information contained in the original bipartite network, and as such typically leads to completely different and not significant outcomes (see Supplementary Note 2).

To sum up, the four main CM-based null models for bipartite network projections proposed in the literature (Hypergeometric, Curveball, BiPCM and BICM) can differ under two aspects. 
The first aspect concerns which constraints are imposed, whether the degrees of one set or both sets of the bipartite network. We can thus speak of ``partial'' models (Hypergeometric and BiPCM) and ``full'' models (Curveball and BiCM). The second aspect concerns how these constraints are imposed,  either exactly (hard constraints) or as ensemble expectations (soft constraints). Using the analogy with statistical physics \cite{cimini2019statistical}, we can refer to these approaches respectively as ``microcanonical'' models (Hypergeometric and Curveball) and ``canonical'' models (BiPCM and BiCM). 
Table \ref{tab:table1} summarizes this classification (see the Methods section for the formal definition of the four null models). 
A fundamental point that has not been addressed so far is whether these formulations lead to different validated networks, 
and thus how to interpret and compare results of the various studies in the literature.

Here we provide, for the first time to our knowledge, a systematic comparison of validation results obtained with the various CM formulations for bipartite network projections. 
We find that albeit based on very similar null hypothesis, the different formulations lead to very different filtered networks even for the same value of validation threshold $p^*$. However we show that a reconciliation of results is possible 
within a region of model-specific thresholds $p^*$ such that the densities of links validated by the null models overlap. In particular we show that a common community structure may emerge in this region. This criterion provides a quantitative approach to build a meta-validated network projection that is independent on the specific implementation of the null model.

\renewcommand{\arraystretch}{2}
\begin{table}[h]
 \caption{Classification of configuration models (CM) for bipartite network projections by number and type of constraints. Partial models only constraints the degrees of nodes belonging to the projection set, while Full models constraints the degrees of nodes in both sets. Hard or microcanonical models impose exact constraints, while Soft or canonical models impose them as ensemble averages. BiPCM stands for Bipartite Partial CM and BICM for Bipartite CM.}
  \label{tab:table1}
  \vspace{0.25cm}
  \begin{tabular}{c|cc}
     & Partial (1 set) & Full (2 sets) \\
  \hline
Hard (microcanonical) & Hypergeometric \cite{teece1994understanding,goldberg2003assessing,tumminello2011statistically} & Curveball \cite{verhelst2008efficient,strona2014fast}\\
Soft (canonical) & BiPCM \cite{saracco2017inferring} & BiCM \cite{gualdi2016statistically,saracco2017inferring}\\    
\end{tabular}
\end{table}

\section*{Results and Discussion}
\label{sec:empirical}

\begin{figure}[h!]
\centering
\includegraphics[width=0.4\textwidth]{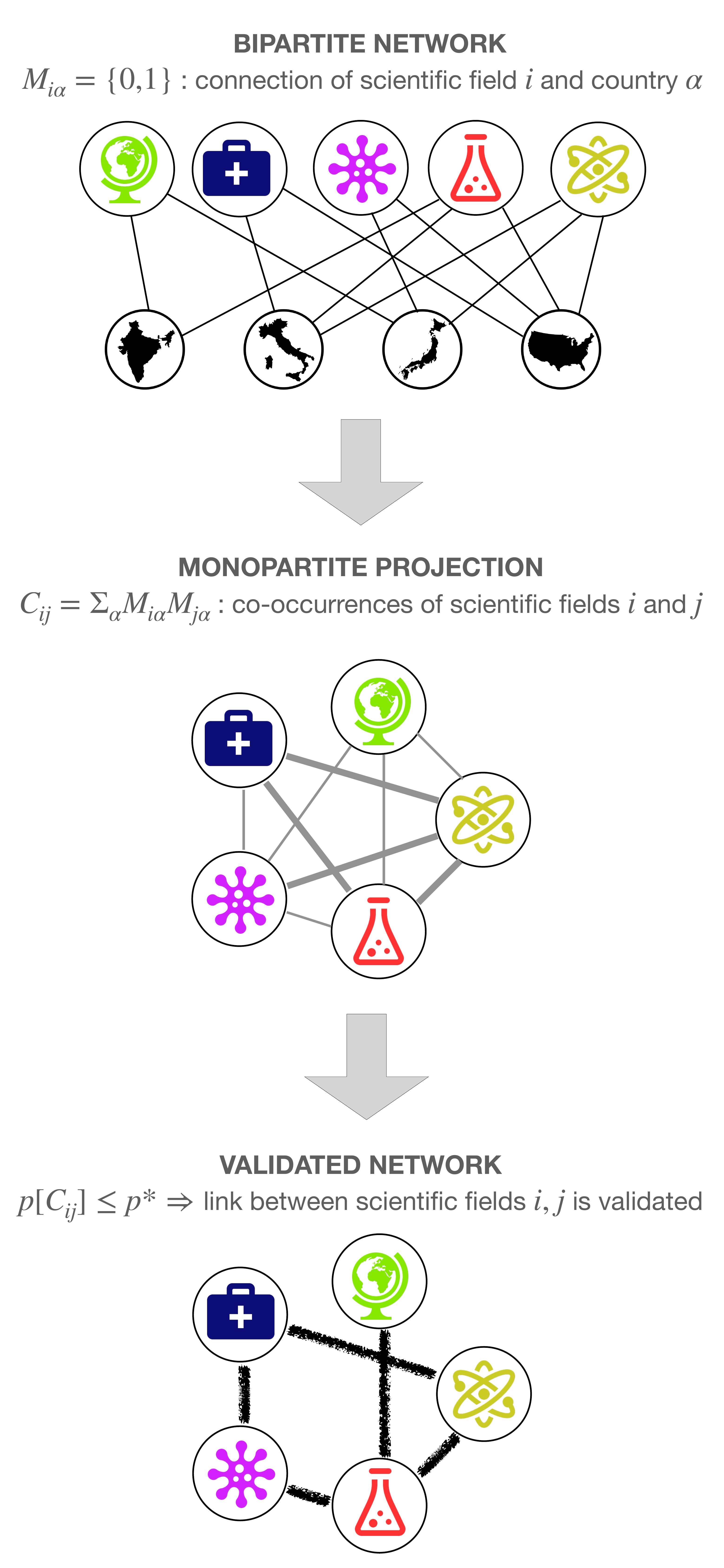}
\caption{{\bf Schematic illustration of the validation procedure for the bipartite network of scientific fields and world countries}. 
We start from the bipartite network $\mx{M}$ of scientific fields 
(in this example: {\em Earth Sciences}, {\em Medicine}, {\em Biology}, {\em Chemistry} and {\em Physics} \cite{IC_sci}) and world countries (here: India, Italy, Japan, United States \cite{IC_cnt}).
In this network, links connect countries with the scientific fields they have a comparative advantage on. 
From this bipartite structure we create a monopartite projected network $\mx{C}$ of scientific fields, whose weighted links represent the co-occurrences of field pairs in the various countries. 
Finally we assess the statistical significance of each observed co-occurrence against its null model expectation: we place a link on the validated network only when the $p$-value is smaller than the significance threshold $p^*$. 
Note that this procedure is general and applies to any bipartite network.}
\label{fig:1}
\end{figure}

We perform the comparison of validation outcomes in the context of co-occurrences for country production networks, following the recent stream of works on economic fitness and complexity \cite{teece1994understanding,hidalgo2007product,klimek2012empirical,zaccaria2014how,pugliese2019unfolding}. 
In particular, our empirical bipartite system is defined by two set of nodes, scientific fields (set $\text{L}$) and world countries (set $\Gamma$), and by links that connect countries with the scientific fields they have a comparative advantage on. 
The $\text{L}$-projection of this bipartite network is a monopartite network of scientific fields, whose generic link $C_{ij}$ is the co-occurrence of fields $i$ and $j$ worldwide. 
Figure \ref{fig:1} summarizes how the validation procedure is applied to this network. 
For a description of raw data and pre-processing, see the Methods section.

We start by recalling the key assumption underlying economic complexity studies on co-occurrences: 
if two scientific fields feature significant co-occurrences (in terms of an appropriate null model) then we can assume that there is an overlap between the capabilities required to achieve proficient level (\ie, competitive advantage) in both fields \cite{pugliese2019unfolding}. 
The need for statistical validation arises in this context since both countries and scientific fields are heterogeneous (if nothing, by their size): two scientific fields may happen to co-occur in many countries just because they are popular worldwide. 
Therefore, a reasonable baseline choice of null model is the (bipartite) CM, for which degrees (\ie, the ubiquity of fields and possibly the diversification of countries) sum up all the information. 
The corresponding null hypothesis is thus that fields are independent and there is no capability structure behind the network: 
co-occurrences between scientific fields happen at random, some more likely than others just because of their ubiquities or countries'  diversification.
Therefore, any specific observed link $i,j$ for which we can reject such null hypothesis is interpreted as the signal of some real interdependence between the specific capabilities required to make proficient scientific research in fields $i$ and $j$.

\subsection*{Different models, different validated networks}

\begin{figure}[p!]
    \begin{minipage}{.475\textwidth}
\centering
\includegraphics[width=0.9\textwidth]{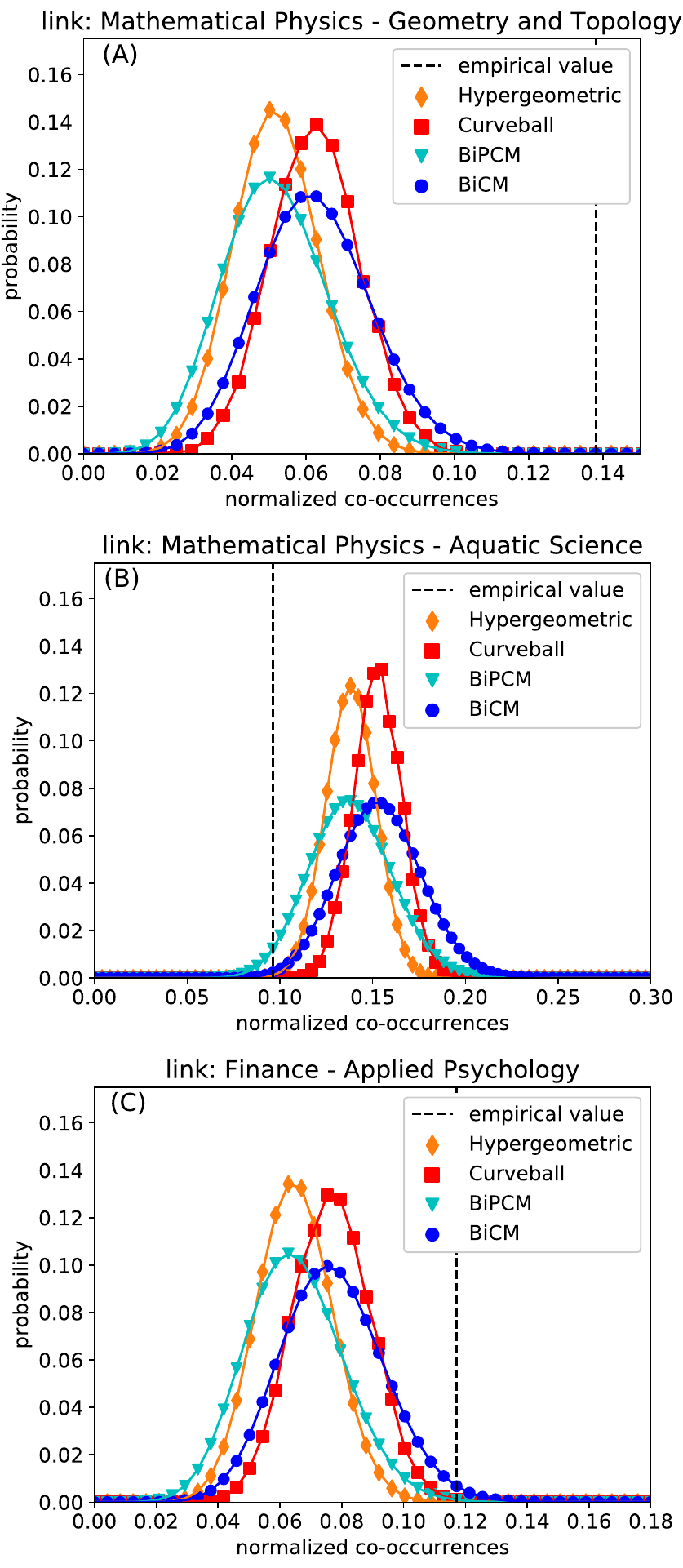}
\caption{{\bf Comparison of empirical co-occurrences and their null model distributions for representative scientific field pairs.} (a) {\em Mathematical Physics - Geometry and Topology}, (b) {\em Mathematical Physics - Aquatic Science},  (c) {\em Finance - Applied Psychology}. For each pair $(i,j)$ of scientific fields we report the empirical value of the (normalized) co-occurrences $C_{ij}/|\Gamma|$ and the respective null model distributions $\pi(\cdot|i,j)$. 
The p-value is given by the area under the distribution starting from the empirical value, hence the outcome of the validation procedure strongly depends both on the significance threshold $p^*$ and the choice of the null model.}
\label{fig:2}
    \end{minipage}%
    \hfill
    \begin{minipage}{0.475\textwidth}
\centering
\includegraphics[width=0.75\textwidth]{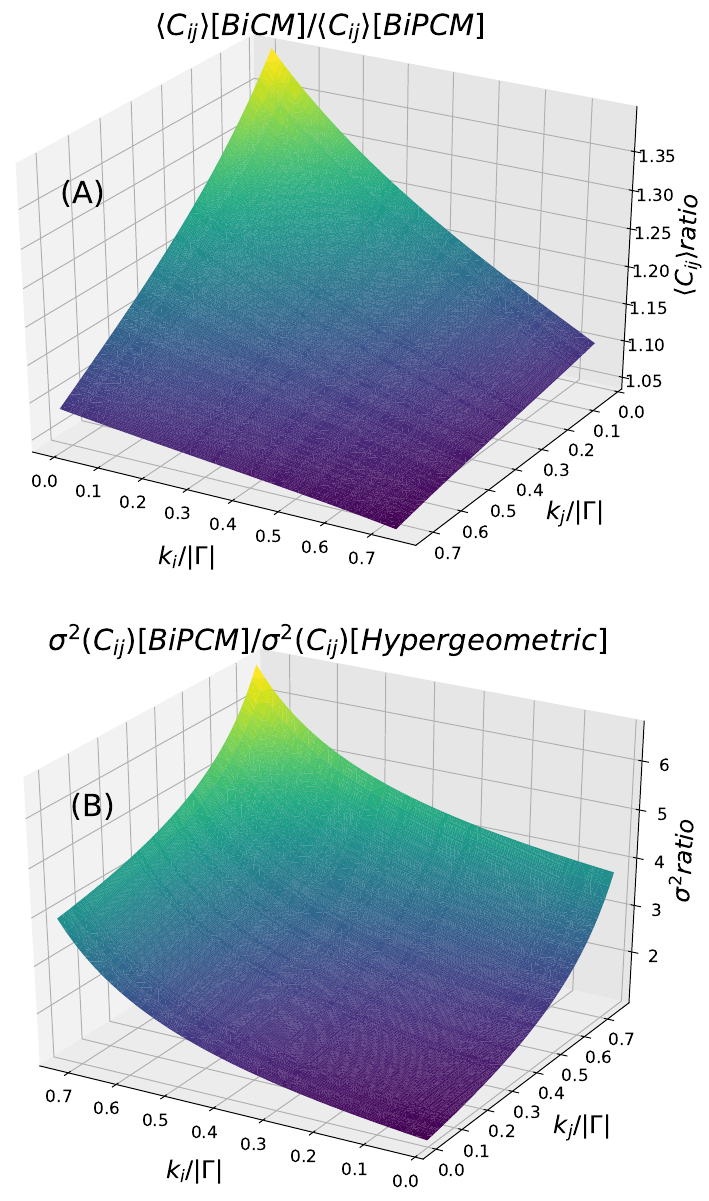}
\caption{{\bf Comparison of null model features}. (a) Ratio of mean co-occurrences $\avg{C_{ij}}$ for BiCM and BiPCM, and (b) ratio of variances $\sigma^2(C_{ij})$ for BiPCM and Hypergeometric, as a function of the normalized degrees $k_i/|\Gamma|$ and $k_j/|\Gamma|$ of the corresponding nodes. Both quantities are strongly dependent on the specific model formulations, especially for high values of the degrees.}  
\label{fig:3}
\bigskip
\includegraphics[width=0.75\textwidth]{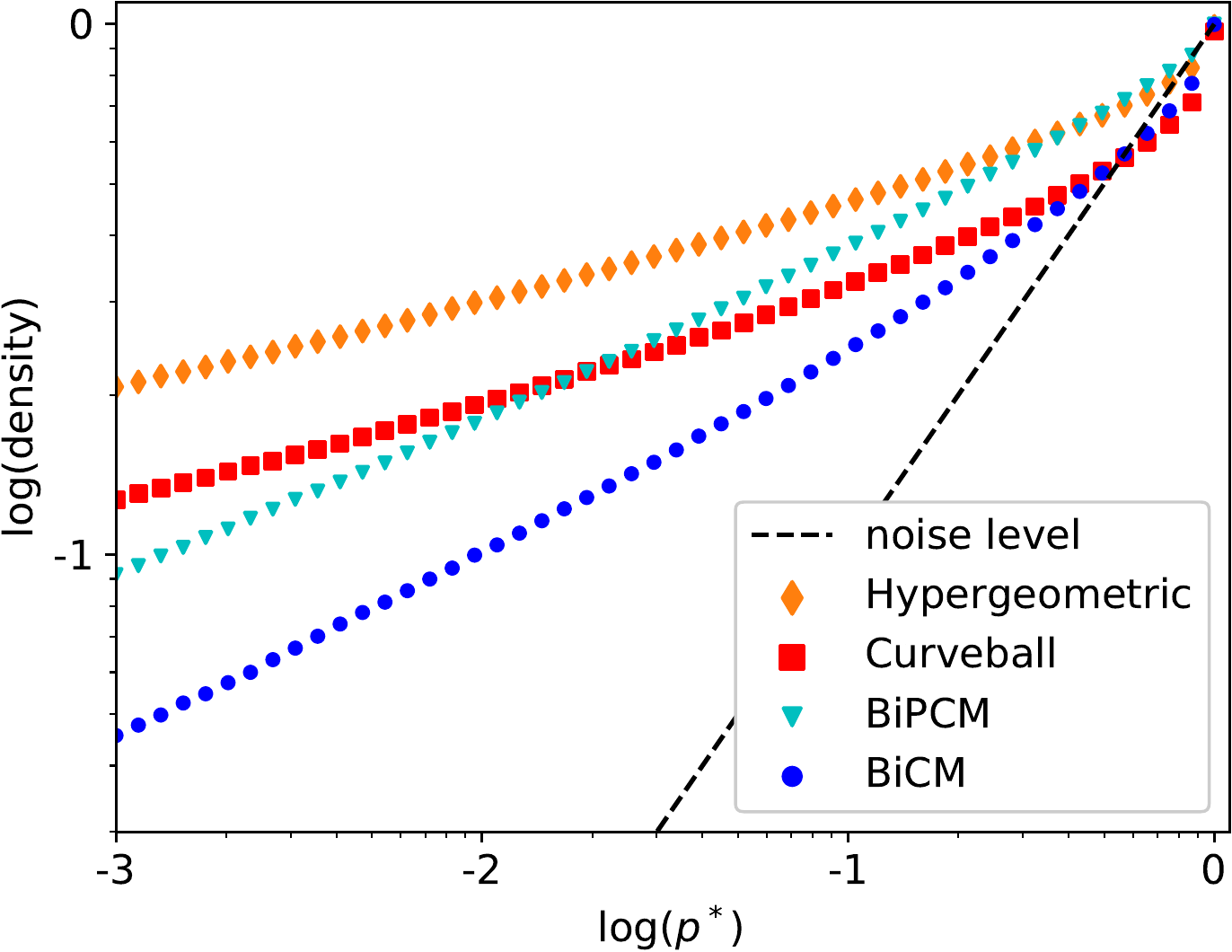}
\caption{{\bf Density $\rho$ of links validated by the various null models as a function of the significance threshold $p^*$}. The dashed bisector denotes the noise level, namely the probability to statistically validate a link generated by the null model. The four null models leads to different results even for the same $p^*$ value.}
\label{fig:4}
    \end{minipage}
\end{figure}

In order to better understand how the validation procedure works, we begin by comparing in Figure \ref{fig:2} the empirical value of the co-occurrences $C_{ij}$ with the respective null model distribution $\pi(\cdot|i,j)$ for some representative pairs $i,j$ of scientific fields. 
We recall that $C_{ij}$ is validated if it satisfies the condition $p[C_{ij}]\le p^*$, that is, if the area under the distribution starting from the empirical value is smaller than the threshold $p^*$. The three example we report are the co-occurrences of: (a) {\em Mathematical Physics - Geometry and Topology}, which are likely validated, in accordance with our expectations that the two fields are related by requiring common skills and capabilities; (b) {\em Mathematical Physics - Aquatic Science}, which are likely not validated, again as we can expect that the two fields are unrelated; and (c) {\em Finance - Applied Psychology}, whose relatedness is plausible but the outcome of the validation procedure is uncertain, as it strongly depends not only on the choice of the threshold $p^*$ but also on the choice of the null model. This is a first evidence that different models may lead to different validated networks. 

These plots also highlight some important symmetries between the various null model distributions (see also Figure \ref{fig:3}). On one hand, the peaks coincide for the two partial models (Hypergeometric and BiPCM), since they have the same average value $\avg{C_{ij}}=k_ik_j/|\Gamma|$ (see eqs. \eqref{avg_MP} and \eqref{avg_CP}), but the same also happens for the two full models (Curveball and BiCM). This means that the average value of the co-occurrences in the null model depends on the set(s) on which degree constraints are imposed. Besides, such average is higher for full models as they capture the heterogeneity of both sets. This can be readily seen by taking the sparse limit of the BiCM, for which eq. \eqref{probfull} becomes $p_{i\alpha}\simeq e^{-\theta_i+t_\alpha}=k_i\kappa_\alpha/E$. Inserting this expression into eq. \eqref{avg_CF} we get $\avg{C_{ij}}\simeq k_ik_j\sum_\alpha \kappa_\alpha^2/E^2$, which is greater than $\avg{C_{ij}}=k_ik_j/|\Gamma|$ of the BiPCM (and equal only when set $\Gamma$ has no heterogeneity, that is, $\kappa_\alpha=E/|\Gamma|$ $\forall \alpha$).
On the other hand, the width of the distribution looks similar for microcanonical models (Hypergeometric and Curveball) and for canonical models (BiPCM and BiCM), implying that the standard deviation of the co-occurrences depends on the types of constraints. As expected, the choice of hard constraints leads to a narrower distribution while the choice of soft constraints leads to a broader distribution. This is easily seen by taking the ratio of variances for the BiPCM (Binomial) and Hypergeometric model, which after some simple algebra can be written as  $(|\Gamma|^2-k_ik_j)/[(|\Gamma|-k_i)(|\Gamma|-k_j)]>1$. Further insights on model comparison are provided in Supplementary Note 3.
Overall, these differences between the null model distributions are likely to produce strong disagreement between the validated networks.

After focusing on individual co-occurrences, we ask in general how many co-occurrences are validated by each null model. We thus measure the link density $\rho(p^*)$ of the validated network, defined as the fraction of $i,j$ pairs such that $p[C_{ij}]\le p^*$, as a function of the significance threshold $p^*$. 
Results for the various null models, reported in Figure \ref{fig:4}, show patterns that are consistent with the above observations. The width of the null model distributions sets the slope of the curves, so that BiCM is the model that validates the least by having longer tails and higher mean, whereas Hypergeometric validates the most by having shorter tails and lower mean. Note also that the Hypergeometric $\rho$ does not tend to 0 for $p^*\to0$: this is due to the distribution not vanishing within its finite support, whereas the effective support of the canonical models distribution is much larger due to the softness of the imposed constraints. Upstream of these considerations, we observe a very large difference in density between the various null models, even of one order of magnitude for the same $p^*$ values. We can thus conclude that the different null models unavoidably lead to different filtered network structures, even when correcting for multiple hypothesis testing (see Supplementary Note 4).

\subsection*{Null models reconciliation}

\begin{figure}[h!]
\centering
\includegraphics[width=0.9\textwidth]{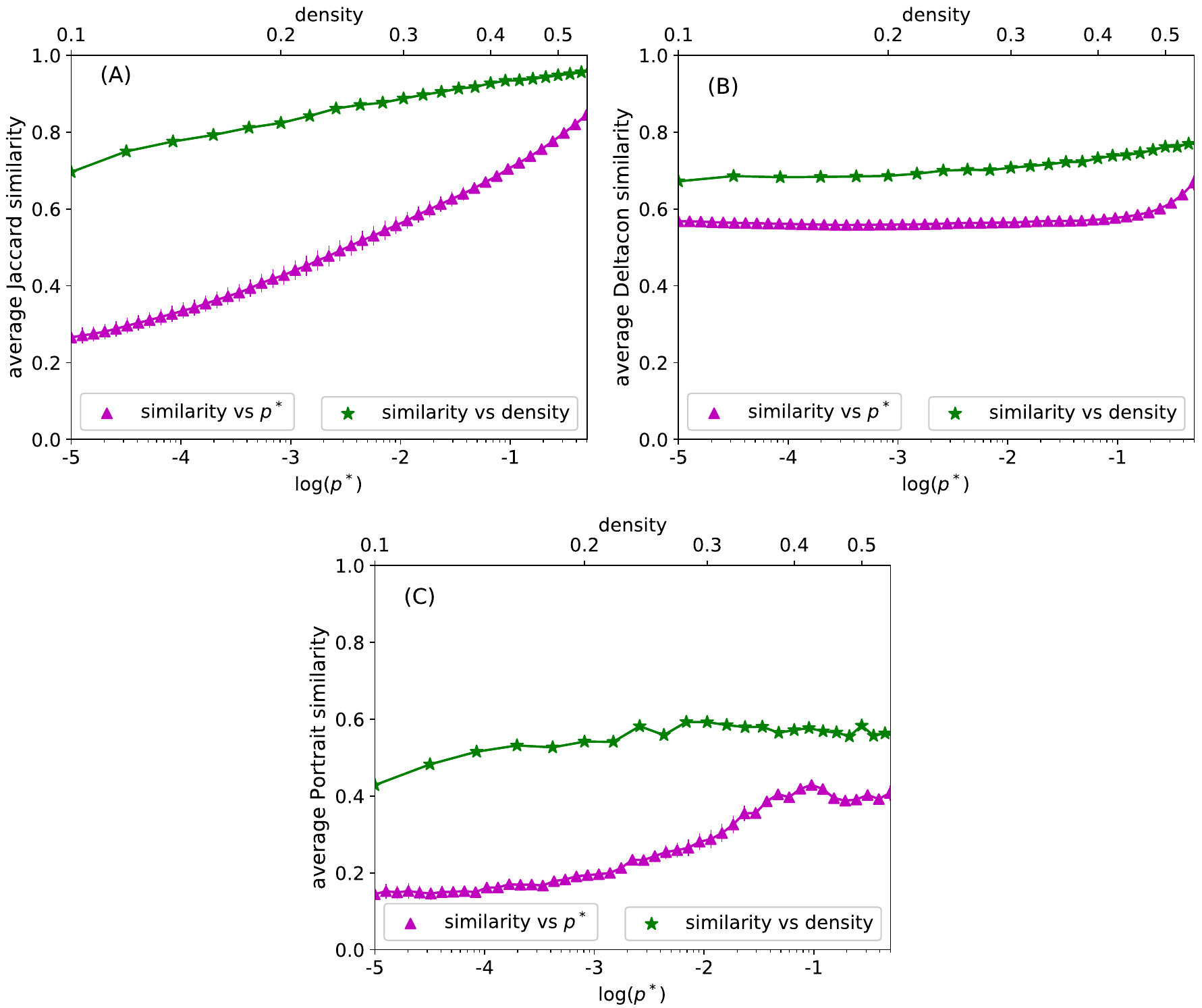}
\caption{ {\bf Average structural similarity of the networks validated by the various null models}. Error bars (not visible) represent standard deviations over choices of null model pairs. Similarity values measured through (a) Jaccard; (b): DeltaCon; (c): Portrait are plotted for filtered networks obtained with the same significance threshold $p^*$ (magenta triangles) or of equal density (green stars). The latter option reveals a higher concordance among the null models. Note how similarity has a baseline dependence on the network density: a change of a link has more impact in lower density graphs.}
\label{fig:5}
\end{figure}

We now discuss a general methodology to reconcile the four validation schemes. The idea is to find a coherence area in the space of parameters where the filtered networks show a relatively good agreement. 
We start by assessing the structural similarity of the validated networks using three popular metrics of graph distance \cite{tantardini2019comparing}. The first one is the simple Jaccard coefficient, which measures the number of links in common between two graphs. Jaccard is a known node-correspondence method, \ie, it requires that the two graphs have the same node set and the pairwise correspondence between nodes is known (our validated networks satisfy this requirement). We further consider: \emph{DeltaCon} \cite{koutra2016delta}, another known node-correspondence method based on the comparison of $l$-length paths connecting each node pair (we use the approximated version of the algorithm, which restricts the computation to randomly chosen pairs); and \emph{Portrait Divergence} \cite{bagrow2019information}, an unknown node-correspondence method that compares the distribution of the shortest-path lengths between graphs. Any of these methods takes as input the adjacency matrices $\mx{V}$ and $\mx{V}'$ of two networks, each validated by a different null model, and returns a measure of their similarity $s_{\mx{VV}'}\in[0,1]$, where $s_{\mx{VV}'}=0$ means the two networks are maximally different while $s_{\mx{VV}'}=1$ that they are identical. 
We can then obtain a mean similarity score by averaging over the six possible choices of null model pairs. A key issue in this comparison is how to choose the validated networks to match. The simplest choice is to compare networks obtained with the same significance threshold $p^*$, and study the average similarity as a function of $p^*$ (Figure \ref{fig:5}, magenta triangles). We see that the average similarity is rather low, especially for Portrait Divergence. In order to recover some compatibility between the results of the different models, we can repeat the operations described above by taking validated networks with the same value of the link density $\rho$ (that is, we adjust $p^*$ for each network in order to obtain the match of $\rho$ values). The resulting curves (green stars in Figure \ref{fig:5}) show that the average similarity of networks at equal $\rho$ is always much higher than for networks at equal $p^*$.

We further study whether the validated  networks are similar in terms of mesoscale or community structure. 
We choose this benchmark because statistical validation on networks is precisely meant to highlight the emergence of multiple-nodes patterns like motifs and communities.
Broadly speaking, a community structure is defined by (typically non-overlapping) sets of nodes, characterized by having many more internal links --- connecting nodes belonging to the same community ---  than external links -- which connect nodes of different communities \cite{fortunato2016community}. In order to find the best partition of the network nodes, a number of community detection algorithms have been proposed in the literature (we remand to  \cite{fortunato2016community} for a recent review of the field). Here we use the popular Louvain method \cite{blondel2008fast}, which is based on maximizing the quality function known as \emph{Modularity} \cite{newman2006modularity}, defined as the observed fraction of links internal to communities with respect to a random benchmark. 
{As there is no community detection method that performs best in all situations \cite{peel2017ground,ghasemian2020evaluating}, in the Supplementary Note 5 we repeat the same analysis using community inference with Bayesian stochastic blockmodeling \cite{peixoto2014efficient} (finding qualitatively similar results).}

\begin{figure}[h!]
\centering
\includegraphics[width=0.75\textwidth]{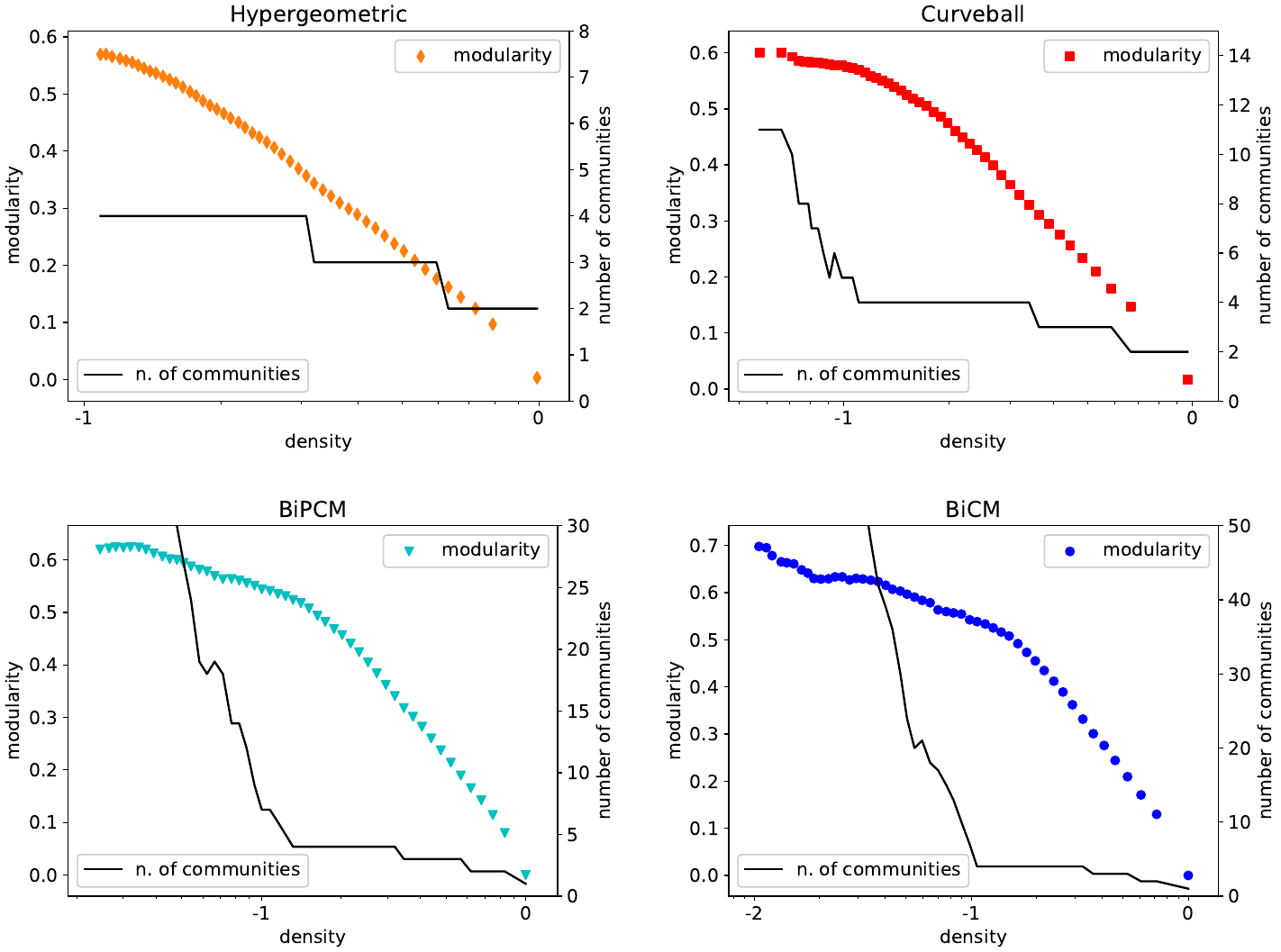}
\caption{{\bf Modularity and number of communities for the best partition obtained by the Louvain method on the validated network, as a function of the density $\rho$ of validated links}. Each point corresponds to the partition of highest modularity obtained in {100} runs of the algorithm with random initialization (hence it has no associated error). While modularity monotonically decreases with the density, we observe that the number of communities has a plateau at 4 for $0.1\lesssim \rho\lesssim 0.3$, suggesting the presence of a region where an agreement between the four approaches is recovered.}
\label{fig:6}
\end{figure}

\begin{figure}[h!]
\centering
\includegraphics[width=0.5\textwidth]{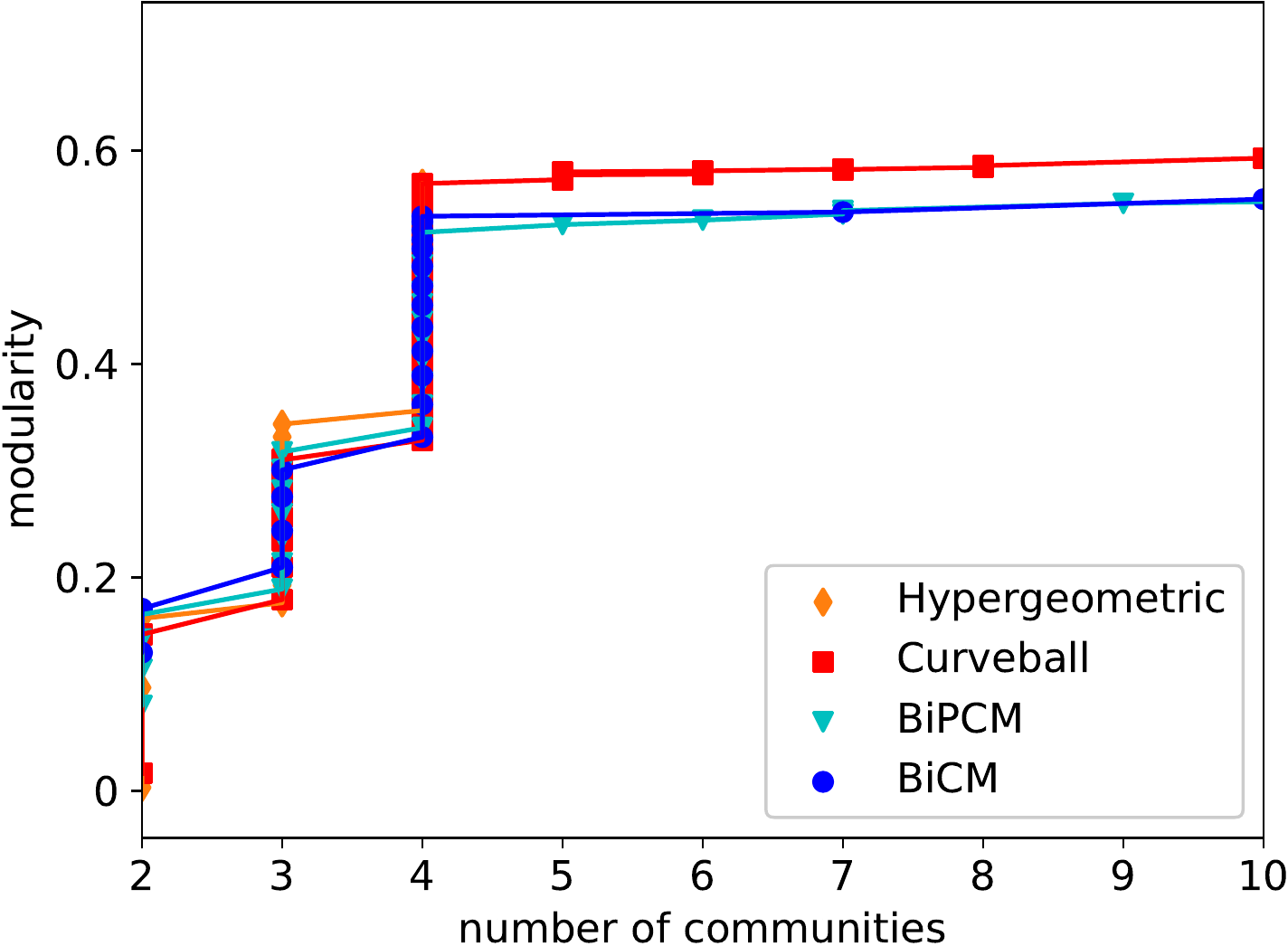}
\caption{{\bf Modularity vs number of communities for the best partition obtained on the validated networks}. Note how the curves of the various models collapse onto each other. Besides, the growth of modularity suddenly stops at 4 communities, suggesting the presence of a robust partition shared between all null models.}
\label{fig:7}
\end{figure}

Figure \ref{fig:6} shows the results of the Louvain algorithm applied on the networks validated by the various null models. Given the previous analysis on structural similarity, we use $\rho$ rather than $p^*$ as independent variable with the aim of achieving a better compatibility between the results of the different models. In each plot, full circles represent the modularity of the best network partition, which increases as the network becomes more sparse, whereas the solid line marks the number of communities, which also increases for decreasing density due to the appearance of more disconnected components. A nontrivial feature that is common to all plots is the presence of a plateau at 4 communities for $0.1\lesssim \rho\lesssim 0.3$. Additionally, modularity tends to stabilize at the lower extreme of this plateau. 
In order to understand whether a community structure common to all null models emerges in this region, we show in Figure \ref{fig:7} the modularity as a function of the number of detected communities.
Using this visualization we get rid of the trivial dependence of modularity and number of communities on $\rho$, and we observe a clear collapse of the curves corresponding to the four models. Additionally we see that modularity, after a fast increase, practically stops to grow after 4 communities are reached (the growth resumes only for a much larger number of communities). This observation points in the direction of a shared community structure. However we still do not know if the partitions identified in each null model setup are actually similar to each other.

\begin{figure}[h!]
\centering
\includegraphics[width=0.5\textwidth]{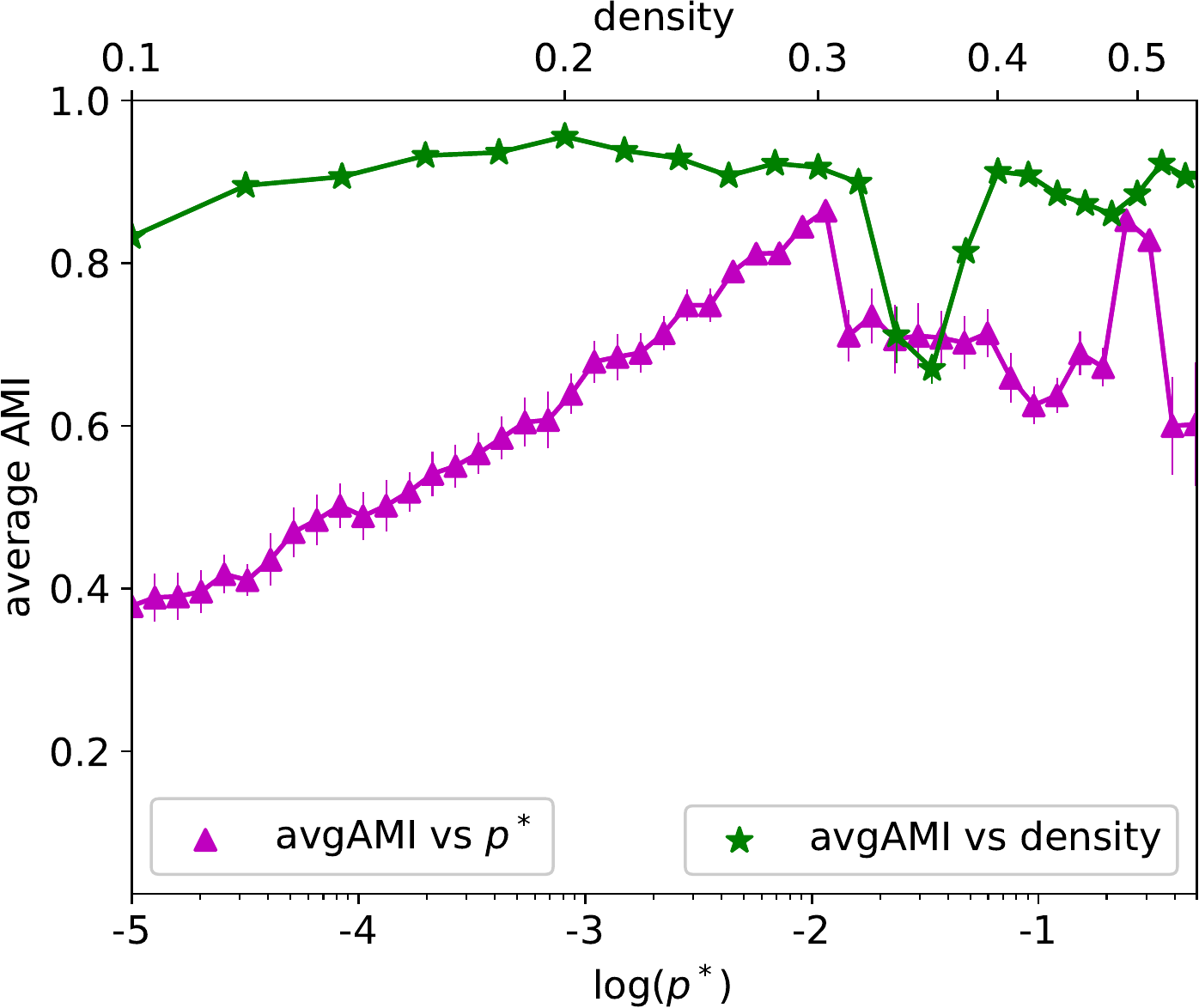}
\caption{{\bf Average Adjusted Mutual Information (AMI) between the best partitions of the network validated by the various null models}. Error bars represent standard deviations over choices of null model pairs. 
Values are plotted for filtered networks obtained with the same significance threshold $p^*$ (magenta triangles) or of equal density (green stars). The latter option reveals a higher concordance among the null models.}
\label{fig:8}
\end{figure}

To quantify the similarity between different partitions we use the \emph{Adjusted Mutual Information} (AMI)  \cite{vinh2010information}. We choose this metrics as it discounts the agreement between partitions solely due to chance, and it is also relatively stable with respect to the presence of disconnected components \cite{romano2014standardized}. Given two network partitions $U_1$ and $U_2$, their AMI is defined as
\begin{equation}
\text{AMI}(U_1,U_2)=\frac {\text{MI}(U_1,U_2)-\mathcal{E}\{\text{MI}(U_1,U_2)\}}{\max\{\text{H}(U_1),\text{H}(U_2)\}-\mathcal{E}\{\text{MI}(U_1,U_2)\}}
\label{eq:ami}
\end{equation}
where $\text{MI}(U_1,U_2)$ is the mutual information between $U_1$ and $U_2$ \cite{strehl2002cluster} while $\text{H}(U_1)$ and $\text{H}(U_2)$ is the Shannon entropy associated with $U_1$ and $U_2$, respectively. 
The adjustment consists in discounting the expected value $\mathcal{E}\{\text{MI}(U_1,U_2)\}$ of the mutual information between two random partitions with the same number of nodes per community as $U_1$ and $U_2$. This correction is needed since the baseline value of mutual information between two random partitions is not constant but grows with the number of communities \cite{vinh2010information}. AMI varies between 0 (if the observed partitions are consistent with a random labelling) and 1 (if the two partitions coincide).

\begin{figure}[h!]
\centering
\includegraphics[width=\textwidth]{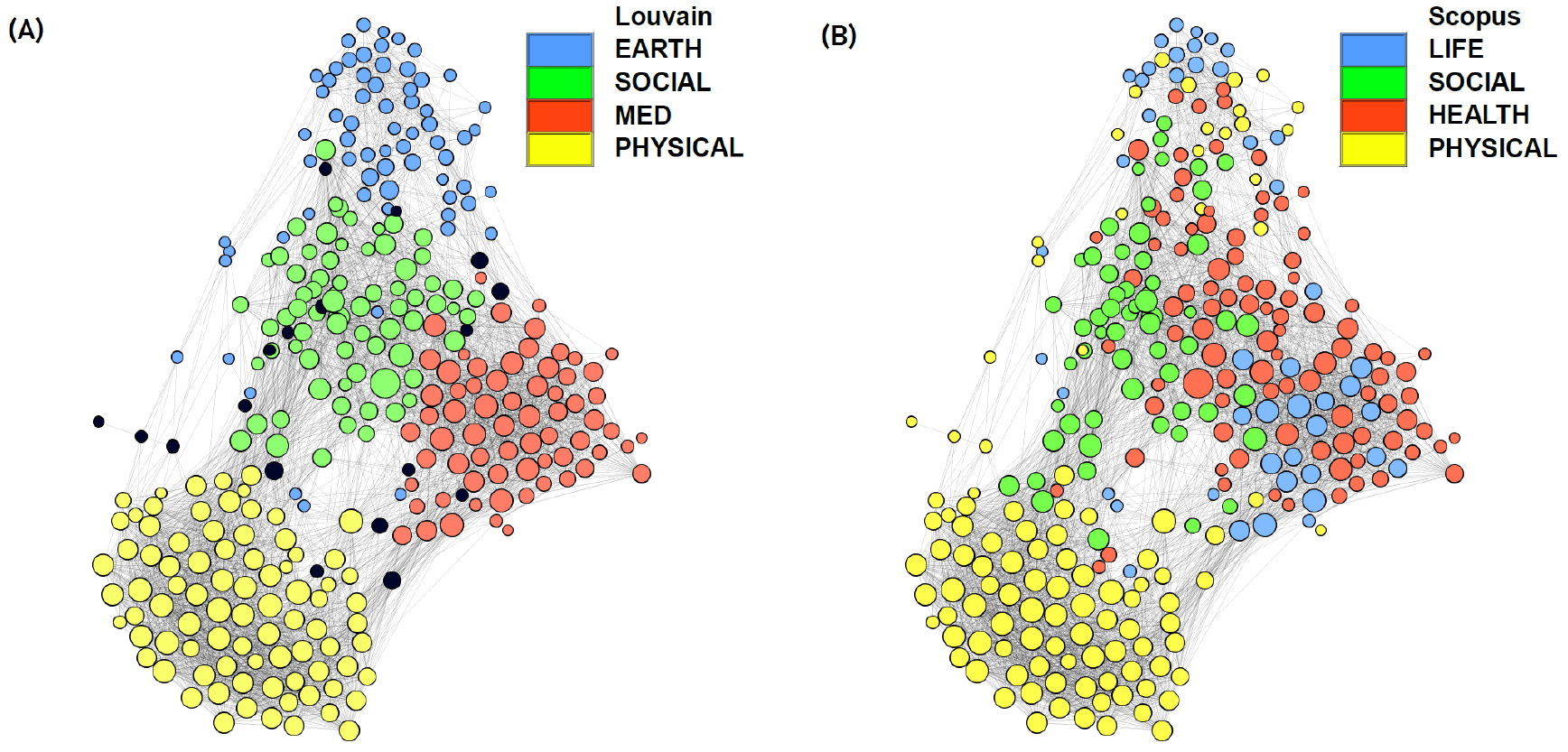}
\caption{ {\bf Community structure of the network of co-occurrence between scientific fields, validated by the Hypergeometric null model.} The link density is $\rho=0.2$, corresponding to the maximum adjusted mutual information (AMI) value of Figure \ref{fig:8}. 
(a) Each color identifies a community (hand-labeled by us as in the legend) shared among the four validated networks, whereas the few black nodes denote the mismatches. The network is represented using a force atlas layout. (b) As a comparison, we report the same network with scientific fields labeled according to their ASJC (All Science Journal Classification) subject area assigned by Scopus' in-house experts. {Visually, the community structure defined by such classification is not much coherent with the network partition induced by our meta-validation approach based on significant co-occurrences, which we recall links two scientific fields when they require common capabilities.}}
\label{fig:9}
\end{figure}

We can thus take a pair of networks each validated using a given null model, extract the respective best partitions (of highest modularity) and compute their AMI. In analogy to what we did for structural similarity metrics, we compare networks that are either validated with the same significance threshold $p^*$, or have the same density (that is, we adjust $p^*$ for each network in order to obtain the match of $\rho$ values). 
We perform this operation for the six possible choices of null model pairs to obtain an average AMI value. Results as a function of $p^*$ or $\rho$ (Figure \ref{fig:8}, magenta triangles or green stars, respectively) confirm that the density $\rho$, and not $p^*$, is the right knob to turn for finding an agreement among the models. Indeed the average AMI computed for networks at equal $\rho$ is almost always higher than AMI for networks at equal $p^*$ --- the only exception being the region around $\rho\simeq 0.3$ where the number of detected communities switches from 4 to 3 but not simultaneously for all models. We can also identify a maximum AMI for $\rho\sim 0.2$, corresponding to the shared community structure illustrated in Figure \ref{fig:9}. Hence the four filtering techniques, which in general produce very different statistically validated networks, can be reconciled by choosing model-specific $p^*$ such that the resulting densities of the validated networks are approximately equal and the AMI is maximum. This strategy is rather general and, in principle, can be applied to any bipartite to monopartite projection in order to produce a ``meta-validated'' filtered network that maximizes the agreement between the different filtering techniques. Even more importantly, it can resolve the arbitrariness in the choice of the significance threshold $p^*$. 

To further support the general applicability of our framework, we show in the {Supplementary Note 6} the same analysis performed to several other bipartite networks belonging to totally different contexts. In all cases we find that both the structural similarity and the AMI of the network partition are consistently higher when computed at equal $\rho$ than when obtained at equal $p^*$. Additionally, when modularity is high enough, a robust community structure shared among the null models emerges.

\section*{Conclusions}
\label{sec:discuss}

The increasing availability of complex data we are experiencing nowadays calls for techniques to extract meaningful information from large-scale networks of interactions. 
The statistical validation of networks is based on comparing empirically observed patterns with their distributional expectation under a null network model. 
This allows performing a statistical test of whether empirical data is explained by the model or represent additional information.
A statistically validated network is built by retaining only the structurally relevant interactions for which the null hypothesis is rejected. This can be of crucial importance to obtain simpler and clearer descriptions of complex systems \cite{cimini2019statistical}. 

For instance, statistical validation of bipartite network projections has been used to detect important patterns in financial markets, such as preferential or avoided relationships \cite{hatzopoulos2015quantifying,musciottoe2021highfrequency}, clusters of investors characterized by the same investment profile \cite{tumminello2012identification,musciotto2016patterns,musciotto2018long}, and overlapping portfolios bearing the highest riskiness for fire sales liquidation \cite{gualdi2016statistically}. 
In the context of economic and innovation systems, validated network projections have been used to detect modules of countries with similar industrial profile 
and the hierarchical structure of products and services \cite{saracco2017inferring,zaccaria2018integrating}, traces of specialisations emerging from the baseline diversification strategy of countries \cite{straka2017grand}, and predictive innovation patterns involving the interplay of scientific, technological and economic activities \cite{pugliese2019unfolding}.
In the context of mobile communications, validated networks were shown to be more resilient than ordinary networks to errors \cite{li2014statistically}. 

Naturally, any null model hinges on a definition of what type of information represents a signal as opposed to noise. As a result, the validated networks obtained through different filtering techniques carry different meanings and highlight different properties. 
Even different constructions of the same null model may yield different outcomes. This latter issue has been recently shown in the context of nestedness in ecological systems \cite{mariani2019nestedness}: whether the degree sequence is responsible for the nestedness of a bipartite network \cite{jonhson2013factors,payrato2019breaking} depends on the choice of the CM-based null model ensemble (microcanonical or canonical) \cite{bruno2020nested}. 
The non-equivalence between the microcanonical and canonical ensembles is due to the extensive number of constraints (one for each network node) and holds also in the thermodynamic limit \cite{barre2007ensemble,anand2009entropy,squartini2015breaking}. 

{In this work we have reviewed (using a unified notation) the CM-based null model formulations for bipartite network projections, and performed a systematic comparison in terms of null model characteristics and validation outcomes within the same contexts.} We showed that the different model formulations lead to different validation results, both at the level of individual links and of macroscale network properties {(see also this recent preprint \cite{neal2021comparing})}. {However we do provide a recipe to reconcile the validation outcomes, by comparing networks obtained with model-specific significance thresholds such that the density of validated links becomes comparable. Additionally this comparison} may allow to identify the region of density values where the agreement between models is maximum. On one hand this solves the arbitrariness in the choice of the significance threshold. On the other hand, it offers a meta-validation approach to identify the filtered configurations with the highest signal-to-noise ratio.

We have included in our comparative study CM-based null models defined on one or both sets of the bipartite network, as well as those defined directly on the monopartite projection, considering in all cases both hard and soft constraints. 
Note that, in principle, ``softer'' constraints could be imposed by fixing the {functional form of the} degree distribution rather than the degree sequence, as in the \emph{hypersoft} CM \cite{anand2014entropy,van2018sparse,voitalov2020weighted}. This approach may be more adequate in the case of dynamic networks, in which degree sequences are never fixed but their  distributions are often stable. Developing a hypersoft CM for bipartite networks and projections, and adding it to our meta-validation framework represents a promising research direction. {Additional challenges for future research are represented by the development of suitable models for weighted bipartite networks and their projections \cite{garlaschelli2009generalized,gabrielli2019grand}, as well as the extension of validation methods beyond pairwise interactions  \cite{battiston2020networks,musciotto2021detecting}.}

As long-term goal we plan to investigate whether the proposed meta-validation approach allows not only to capture the most relevant structural properties of the network projection, but also to help in predicting its evolution -- namely, which links will appear in the future. This could be important in various contexts, from link prediction for recommender systems based on collaborative filtering \cite{zhou2007bipartite,kobayashi2019structure} to assessing prices trend and systemic risk in financial networks of portfolios and assets \cite{gualdi2016statistically,vodenska2021systemic} and forecasting development patterns in economic and innovation systems \cite{pugliese2019unfolding,tacchella2021relatedness,straccamore2021firms}.

\section*{Methods}

\subsection*{Null Models of Bipartite Network Projections}

Here we provide the mathematical definitions of the null models used in our analysis.

\paragraph*{Microcanonical Partial Model:  Hypergeometric.}
For the projection of a bipartite network on set $\text{L}$, an analytic null hypothesis can be formulated by assuming random connections between nodes of the two sets $\text{L}$ and $\Gamma$ that preserve the degree heterogeneity of set $\text{L}$ \cite{teece1994understanding,tumminello2011statistically}. 
Under this hypothesis, the probability that nodes $i$ and $j$ have $x$ co-occurrences is given by the hypergeometric distribution
\begin{equation}\label{pi_MP}
\pi(x|i,j)=\binom{k_i}{x}\binom{|\Gamma|-k_i}{k_j-x}\Big/\binom{|\Gamma|}{k_j}
\end{equation}
and the mean value of the co-occurrences is
\begin{equation}\label{avg_MP}
\avg{C_{ij}}=k_ik_j \Big/|\Gamma|.
\end{equation}
This probability is exact only when nodes of set $\Gamma$ have the same degree. A tentative extension to deal with the degree heterogeneity of set $\Gamma$ consists in splitting the original bipartite network into subnetworks each consisting of set $\Gamma$ nodes with the same degree and of all set $\text{L}$ nodes linked to them, so that the null hypothesis can be properly cast for each subnetwork \cite{tumminello2011statistically}. However, when set $\Gamma$ is highly heterogeneous these subnetworks are many in number and very sparse, causing severe resolution issues (see the discussion in \cite{gualdi2016statistically}).

\paragraph*{Microcanonical Full model: Curveball.}
Building a null bipartite network model where only configurations with a given degree sequence for both sets of nodes are allowed has not been tackled analytically up to now {(exact results exist only in the thermodynamic limit concerning the count of bipartite graphs with given degree sequences \cite{liebenau2020asymptotic})}. This model is hard to deal with because, differently from the canonical case (see below), link probabilities are not pairwise independent. Therefore the model must be defined through an ensemble of bipartite network configurations that are generated numerically by swapping links iteratively so to preserve degrees exactly. 
The \emph{Curveball} algorithm \cite{verhelst2008efficient,strona2014fast,carstens2015proof} works as follows: Starting from the empirical bipartite network $\mx{M}$, these steps are repeated $n$ times:
\begin{enumerate}
    \item Select at random a pair of nodes $i,j$ in set $\text{L}$;
    \item Check that the neighborhoods of the nodes are not perfectly overlapping (otherwise start again);
    \item Take the set of uncommon neighbors $\delta(i,j) =\{\alpha\in \Gamma\,|\,M_{i\alpha}\oplus M_{j\alpha}=1\}$ and remove them from the neighborhood of both;
    \item Assign $k_i - \sum_\alpha M_{i\alpha}M_{j\alpha}$ new neighbors to node $i$, chosen at random from $\delta(i, j)$, and the rest of the nodes in $\delta(i, j)$ to node $j$.
\end{enumerate}
The result is a randomized bipartite network configuration $\mxt{M}$ (here and in what follows we use the tilde symbol to denote matrix configurations of the null model). This procedure is repeated iteratively to generate an ensemble $\{\mxt{M}_q\}_{q=1}^Q$ of $Q$ independent randomizations of the bipartite network. The null model ensemble $\{\mxt{C}_r\}_{r=1}^R$ of projected networks is then obtained by projecting pairs of different instances of bipartite randomizations (that is, a generic configuration is obtained as $\mxt{C}_r=\mxt{M}_q\mxt{M}^T_{q'}$ with $q\neq q'$). The null model distributions $\pi(\cdot|i,j)$ $\forall i,j$ are then computed numerically by sampling from such an ensemble.
Here we use $n=5\min\{|\text{L}|,|\Gamma|\}$ and $Q=R=10000$. 
Note that for a numerically-generated ensemble of $R$ network configurations, the minimum p-value that can be used for statistical testing is $1/R$.

\paragraph*{Canonical Full model: BiCM.}
Generally speaking, canonical models of networks \cite{park2004statistical,bianconi2008entropy,garlaschelli2008maximum,squartini2011analytical} (also known as \emph{exponential random graphs} 
\cite{holland1981exponential,strauss1986general,snijders2006new}) define an ensemble $\Omega$ of networks using a constrained entropy maximization procedure, which leads to assuming the utmost ignorance about the unconstrained degrees of freedom of the system \cite{jaynes1957information,cimini2019statistical}. 
The Bipartite Configuration Model (BiCM) \cite{saracco2015randomizing} applies to bipartite networks by constraining the ensemble average of the degree sequence for both node sets.
The ensemble probability distribution that maximizes the Shannon entropy under these constraints is 
\begin{equation}
P(\mxt{M}|\{\theta_i\},\{t_\alpha\})=e^{-H(\mxt{M},\{\theta_i\},\{t_\alpha\})}/Z(\{\theta_i\},\{t_\alpha\})\label{eq:prob}
\end{equation}
where $\{\theta_i\}$ and $\{t_\alpha\}$ are the sets of Lagrange multipliers associated to the constraints $\{k_i\}$ and $\{\kappa_\alpha\}$ respectively, $Z(\{\theta_i\},\{t_\alpha\})=\sum_{\mx{M}\in\Omega} e^{-H(\mxt{M},\{\theta_i\},\{t_\alpha\})}$ is the partition function and the Hamiltonian $H(\mxt{M},\{\theta_i\},\{t_\alpha\}) =\sum_{i\in\text{L}}\theta_i k_i(\mxt{M})+ \sum_{\alpha\in\Gamma} t_\alpha \kappa_\alpha(\mxt{M})$ sums up the imposed constraints. 
Note that $P(\mxt{M}|\{\theta_i\},\{t_\alpha\})$ depends on $\mxt{M}$ only through $k_i(\mxt{M})$ and $\kappa_\alpha (\mxt{M})$: network configurations with the same value of the constraints are equiprobable, which implies that the canonical ensemble is maximally non-committal {(or the least biased)} with respect to the properties that are not enforced on the system. 
Since degrees are linear constraints the partition function can be computed analytically, so the ensemble probability factorizes as
\begin{equation}
P(\mxt{M}|\{\theta_i\},\{t_\alpha\})=\prod_{i, \alpha}p_{i\alpha}^{\tilde{M}_{i\alpha}}(1-p_{i\alpha})^{1-\tilde{M}_{i\alpha}}
\end{equation}
where $p_{i\alpha}$ is the existence probability of the link connecting nodes $i$ and $\alpha$:
\begin{equation}
p_{i\alpha}=(e^{\theta_i+t_\alpha}+1)^{-1}.
\label{probfull}
\end{equation}
The numerical values of the link probabilities (\ie, of the Lagrange multipliers) are determined by maximising the likelihood of the empirical bipartite network $\mx{M}$ in the ensemble, which implies solving the constraints equations
\begin{equation}\label{constraints}
\left\{ 
\begin{array}{ll}
k_i=\sum_{\alpha\in\Gamma} p_{i\alpha} &\quad i\in\text{L}\\
\kappa_\alpha=\sum_{i\in\text{L}} p_{i\alpha} &\quad \alpha\in\Gamma
\end{array}
\right.
\end{equation}
Once link probabilities have been found, the expected co-occurrences between any two nodes $i \neq j$ are
\begin{equation}
 \avg{C_{ij}}=\sum_{\alpha\in\Gamma}p_{i\alpha}p_{j\alpha},
 \label{avg_CF}
\end{equation}
and the probability distribution $\pi(\cdot|i,j)$ of this quantity is the distribution of the sum of $\Gamma$ independent Bernoulli trials, each with success probability $p_{i\alpha}p_{j\alpha}$. 
This is a Poisson-Binomial distribution, which can be computed numerically \cite{gualdi2016statistically} or analytically \cite{saracco2017inferring} as 
\begin{equation}
\label{pi_CF}
\pi(x|i,j)=\sum_{\gamma_x}\left[\prod_{\alpha\in\gamma_x} p_{i\alpha}p_{j\alpha}\prod_{\beta\notin\gamma_x}(1-p_{i\beta}p_{j\beta})\right]
\end{equation}
where $\gamma_x$ denotes all possible $x$-tuples of nodes in set $\Gamma$.

\paragraph*{Canonical Partial model: BiPCM.}
The ``partial'' version of the BiCM, named BiPCM in ref. \cite{saracco2017inferring}, is defined as the canonical model that constrains only the degree sequence of set $\text{L}$. As such, it is a special case of the BICM described above where all Lagrange multipliers $\{t_\alpha\}$ associated with degrees of set $\Gamma$ are ``switched off'' (\ie, set equal to zero). The Hamiltonian is thus $H(\mxt{M},\{\theta_i\}) =\sum_{i\in\text{L}}\theta_i k_i(\mxt{M})$ and the link probability of generic link $(i,\alpha)$ becomes $p_{i\alpha}=(e^{\theta_i}+1)^{-1}$. Using the constraint equations
$k_i=\sum_{\alpha\in\Gamma} p_{i\alpha}$ $\forall i\in\text{L}$ we get
the explicit expression
\begin{equation}
p_{i\alpha}=k_i\Big /|\Gamma|\qquad\forall i\in\text{L}.
\end{equation}
Therefore the expected value of the co-occurrence between any two nodes $i$ and $j$ is
\begin{equation}
 \avg{C_{ij}}=k_ik_j\Big /|\Gamma|
 \label{avg_CP}
\end{equation}
and its distribution has a simple Binomial form
\begin{equation}
\label{pi_CP}
\pi(x|i,j)=\binom{|\Gamma|}{x}\left(\frac{k_ik_j}{|\Gamma|^2}\right)^{x}\left(1-\frac{k_ik_j}{|\Gamma|^2}\right)^{|\Gamma|-x}
\end{equation}

\subsection*{Data, RCA filter and Projection}
\label{appData}

To build the bipartite network of countries and scientific fields, we use data on scientific productivity and impact of countries collected from the SCIMAGO platform (based on Scopus). 
The database contains the corpus of scientific publications in journals, book series, conference proceedings, and books in the various scientific fields, covering the time interval from $1996$ to $2018$. 
Data are then aggregated at the level of countries and scientific fields (in total there are $|\text{L}| = 307$ scientific fields and $|\Gamma| = 239$ countries), so that $W_{i\alpha}$ is the total number of scientific documents produced by country $\alpha$ in scientific field $i$ during the time span of the data.

In order to determine whether a given country $\alpha$ shows a comparative advantage in field $i$, both with respect to other countries as well as to other fields, the \emph{revealed comparative advantage} (RCA) \cite{balassa1965trade} filter comes at hand. While originally developed in the economic context, this metric has also found use in studies of scientific production \cite{bowen1983thepretical,guevara2016research,pugliese2019unfolding}. 
RCA is an intensive indicator computed as the ratio between the weight of field $i$ in the scientific basket of country $\alpha$ and the weight of field $i$ in the total world science. As a comparative advantage is revealed if $\text{RCA}>1$, we binarize the raw matrices to obtain new matrices
\begin{equation}
  M_{i\alpha}=
  \left\{\begin{array}{ll}
        1\;\mbox{ if }\dfrac{W_{i\alpha}}{\sum_jW_{j\alpha}}\Bigg/\dfrac{\sum_{\beta} W_{i\beta}}{\sum_{j\beta}W_{j\beta}}\ge1,\\
        0\;\mbox{ otherwise. } 
        \end{array}
  \right.
\end{equation}
Note that the RCA filter is properly normalized by making quantities related to different countries and fields comparable \cite{radicchi2008universality}.

Once the binary bipartite matrix is defined, we build the projected network of co-occurrences between scientific fields, whose generic connection between fields $i$ and $j$ is $C_{ij}=\sum_{\alpha}M_{i\alpha}M_{j\alpha}$. 
Note that for the sake of having a a clearer picture and more analytical insights on the various null models, we do not employ here more refined formulations of co-occurrences that use additional normalization by degrees \cite{hidalgo2007product,zaccaria2014how,pugliese2019unfolding}.

\bigskip

\paragraph*{Acknowledgments.} 
We thank Benedetta Castagna and Aurelio Patelli for useful discussion. 
We acknowledge the CREF project "Complessità in Economia" and the ISC-CNR project "CompLang".

\bigskip

\paragraph*{Author contributions.}
GC and AZ designed the research. AC and LD performed research. AC realised the figures. GC wrote the manuscript. GC, AC and AZ reviewed the manuscript.

\bigskip

\paragraph*{Data availability.}
The dataset about scientific productivity of countries analysed during the current study can be obtained from SCImago, (n.d.). SJR - SCImago Journal \& Country Rank [Portal], which can be retrieved at \url{https://www.scimagojr.com/countryrank.php}. The other datasets analyzed in the Supplementary Note 6 are available at the web addresses indicated in the same document.

\bigskip

\paragraph*{Code availability.}
The code to run the Curveball algorithm can be retrieved from \cite{strona2014fast}, while the code to run BiCM is available at \url{https://github.com/tsakim/bicm} (Hypergeometric and BiPCM have analytic formulas). Codes for computing network distance metrics can be retrieved from \cite{tantardini2019comparing} while the code for computing Modularity and AMI are available respectively at \url{https://github.com/taynaud/python-louvain} and \url{https://scikit-learn.org/stable/modules/clustering.html#mutual-info-score} \cite{JMLR:v12:pedregosa11a}. Network visualizations have been generated using Gephi \url{https://gephi.org}.

\bigskip




\section*{Supplementary material}

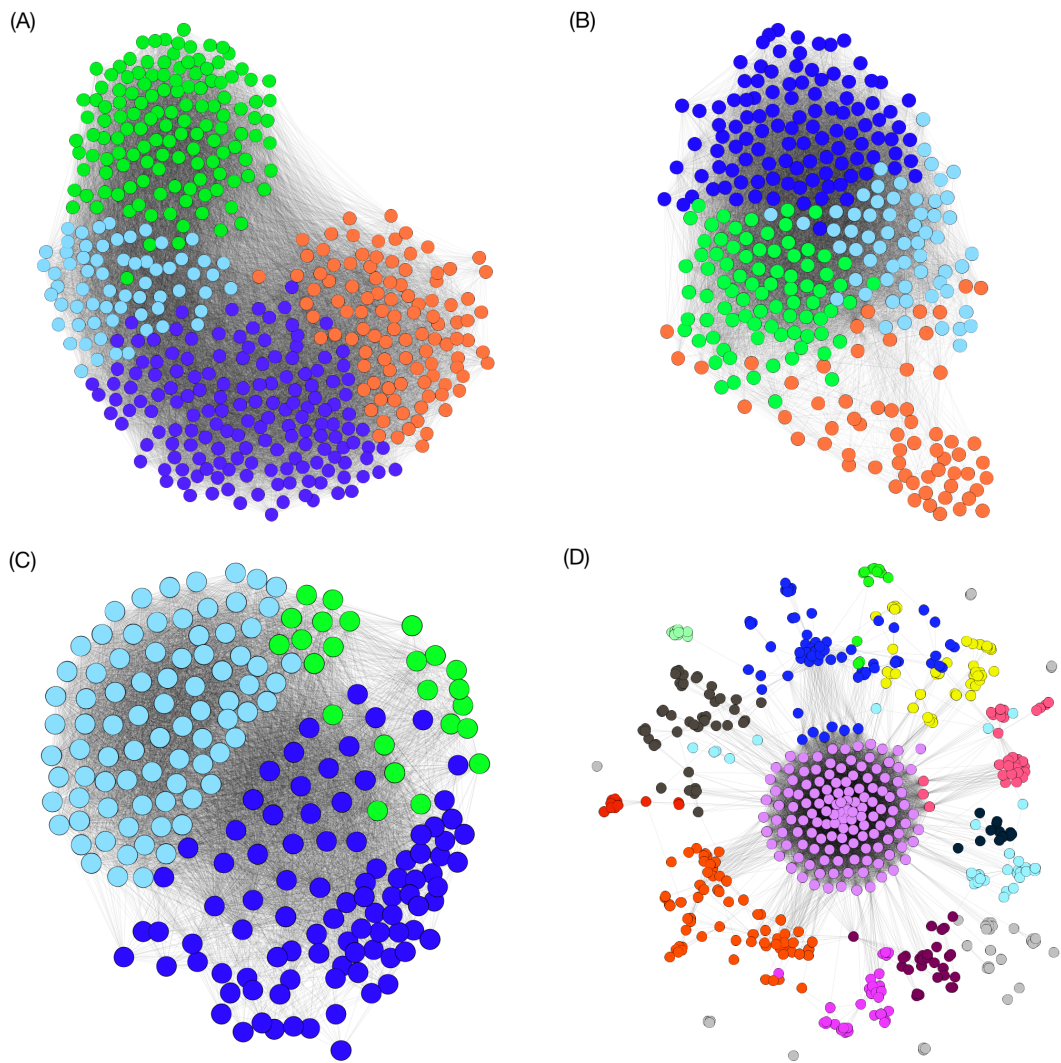

Supplementary Figure 12. Community structures of other networks in our dataset validated by the BiCM null model. Each color identifies a community shared among the four validated networks. (A) *ROBERT* network of co-occurrence between plant species, defined in terms of animal species they are visited by, for $\rho = 0.29$ corresponding to maximum AMI as of Supplementary Figure 14, bottom right panel. (B) *TAYLOR* network of co-occurrence between cities, defined in terms of the different large firms office they host in common, for $\rho = 0.36$ corresponding to maximum AMI as of Supplementary Figure 15, bottom right panel. (C) *DUTCH* network of co-occurrence between corporate elites, defined in terms of the different boards they both sit on, for $\rho = 0.43$ corresponding to maximum AMI as of Supplementary Figure 16, bottom right panel. (D) *CRIME* network of co-occurrence between persons, defined in terms of the different cases they appeared in, for $\rho = 0.08$ corresponding to maximum AMI as of Supplementary Figure 17, bottom right panel.

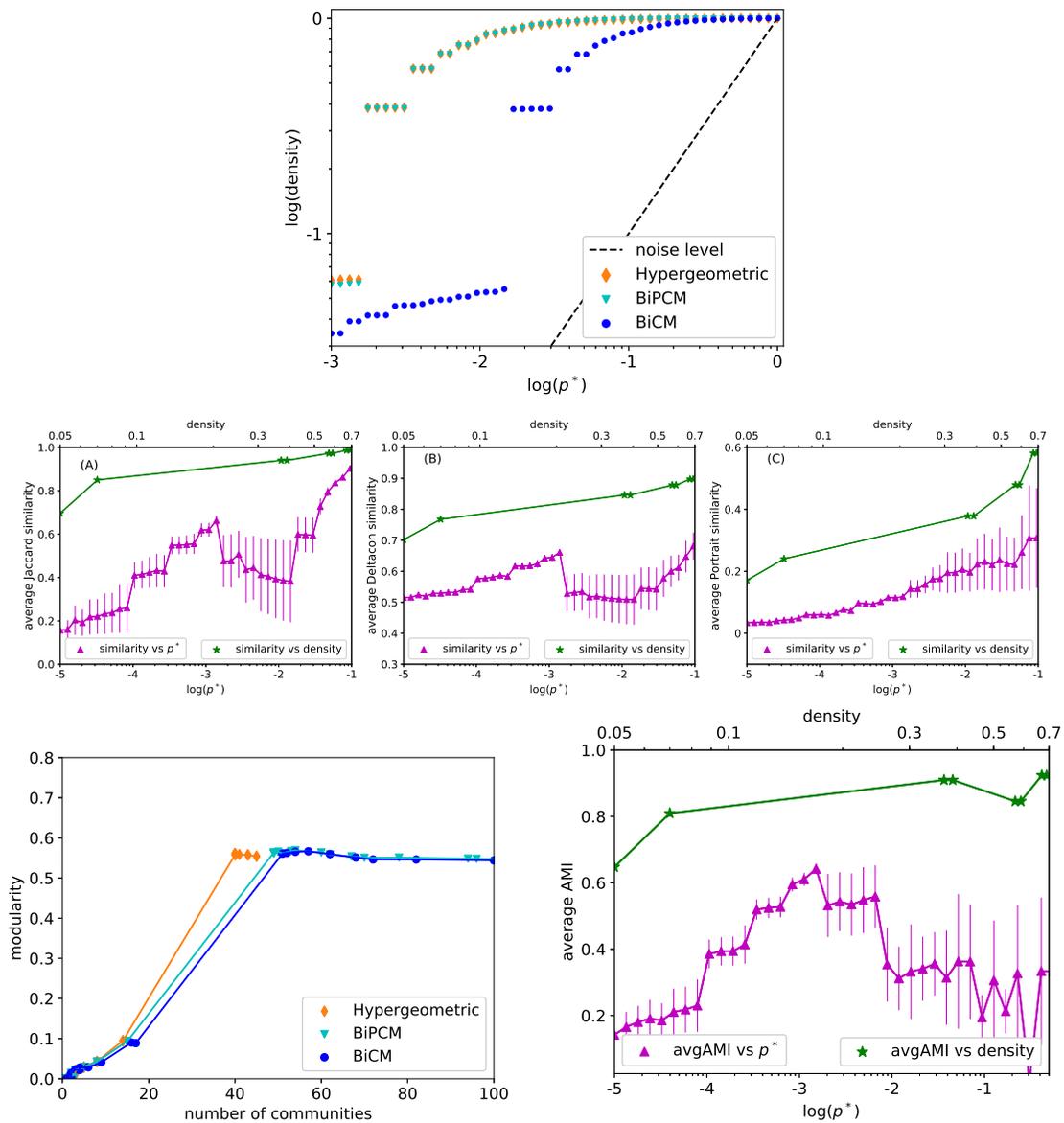

Supplementary Figure 13. *HOST-VIRUS* network. Top panel: density $\rho$ of validated links as a function of $p^*$. Middle panels: (A) Jaccard, (B) DeltaCon, (C) Portrait similarity of validated networks, as a function of either $p^*$ or $\rho$. Bottom panels: (left) Modularity vs number of communities; (right) average AMI for partitions of the validated networks as a function of either $p^*$ or $\rho$. Error bars represent standard deviations over choices of null model pairs.

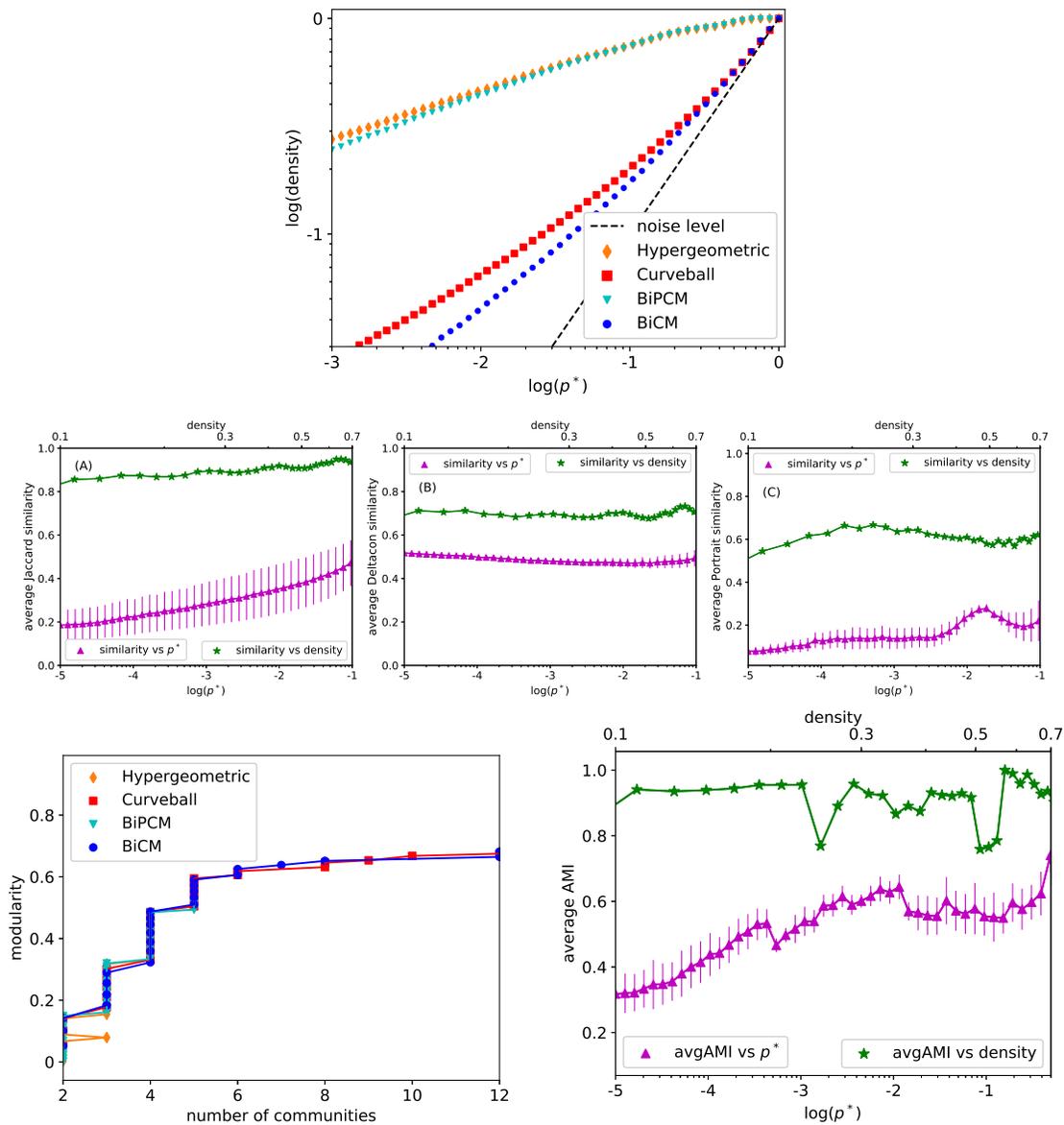

Supplementary Figure 14. *ROBERT* network. Top panel: density $\rho$ of validated links as a function of $p^*$. Middle panels: (A) Jaccard, (B) DeltaCon, (C) Portrait similarity of validated networks, as a function of either $p^*$ or $\rho$. Bottom panels: (left) Modularity vs number of communities; (right) average AMI for partitions of the validated networks as a function of either $p^*$ or $\rho$. Error bars represent standard deviations over choices of null model pairs.

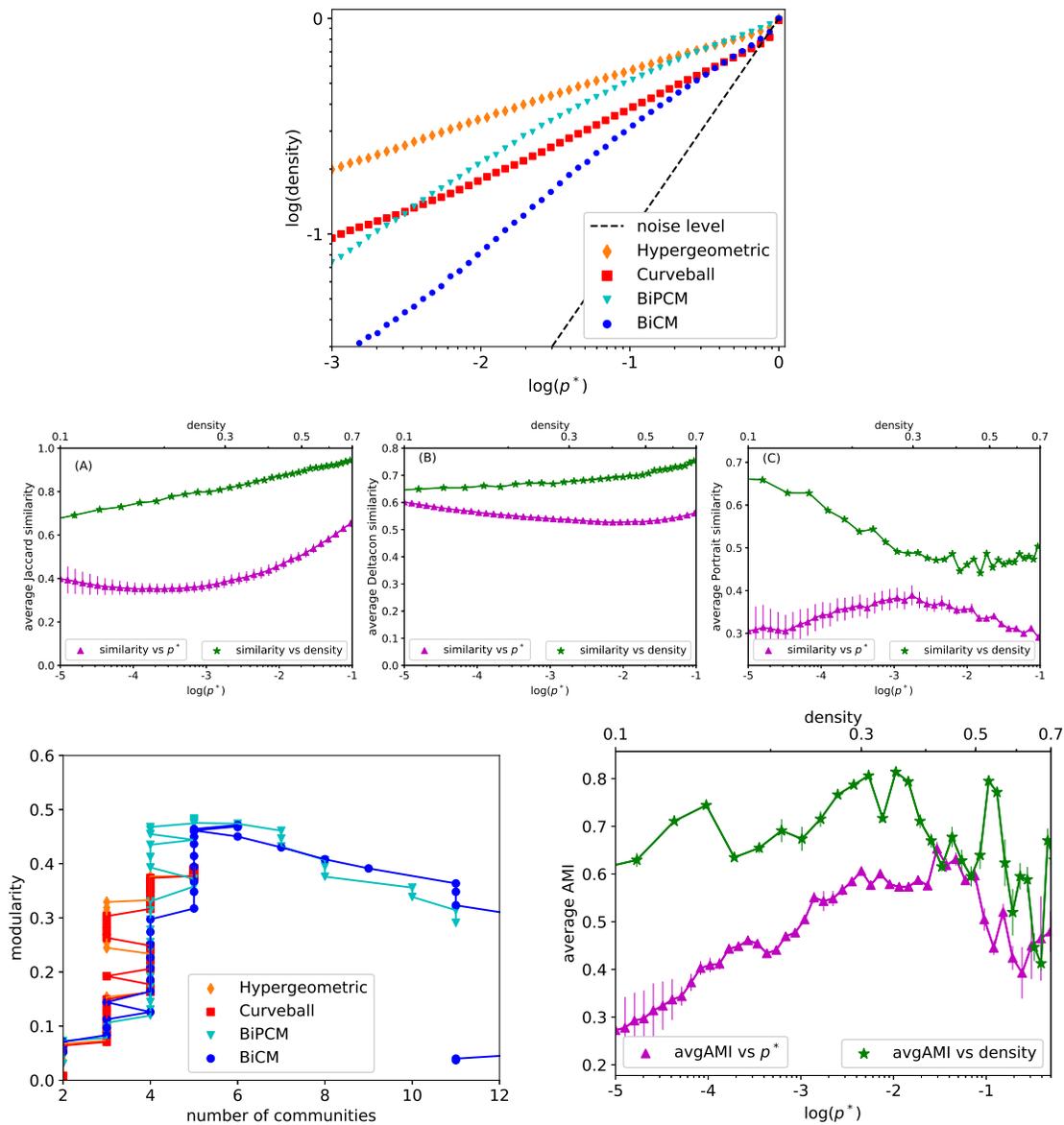

Supplementary Figure 15. *TAYLOR* network. Top panel: density $\rho$ of validated links as a function of $p^*$. Middle panels: (A) Jaccard, (B) DeltaCon, (C) Portrait similarity of validated networks, as a function of either $p^*$ or $\rho$. Bottom panels: (left) Modularity vs number of communities; (right) average AMI for partitions of the validated networks as a function of either $p^*$ or $\rho$. Error bars represent standard deviations over choices of null model pairs.

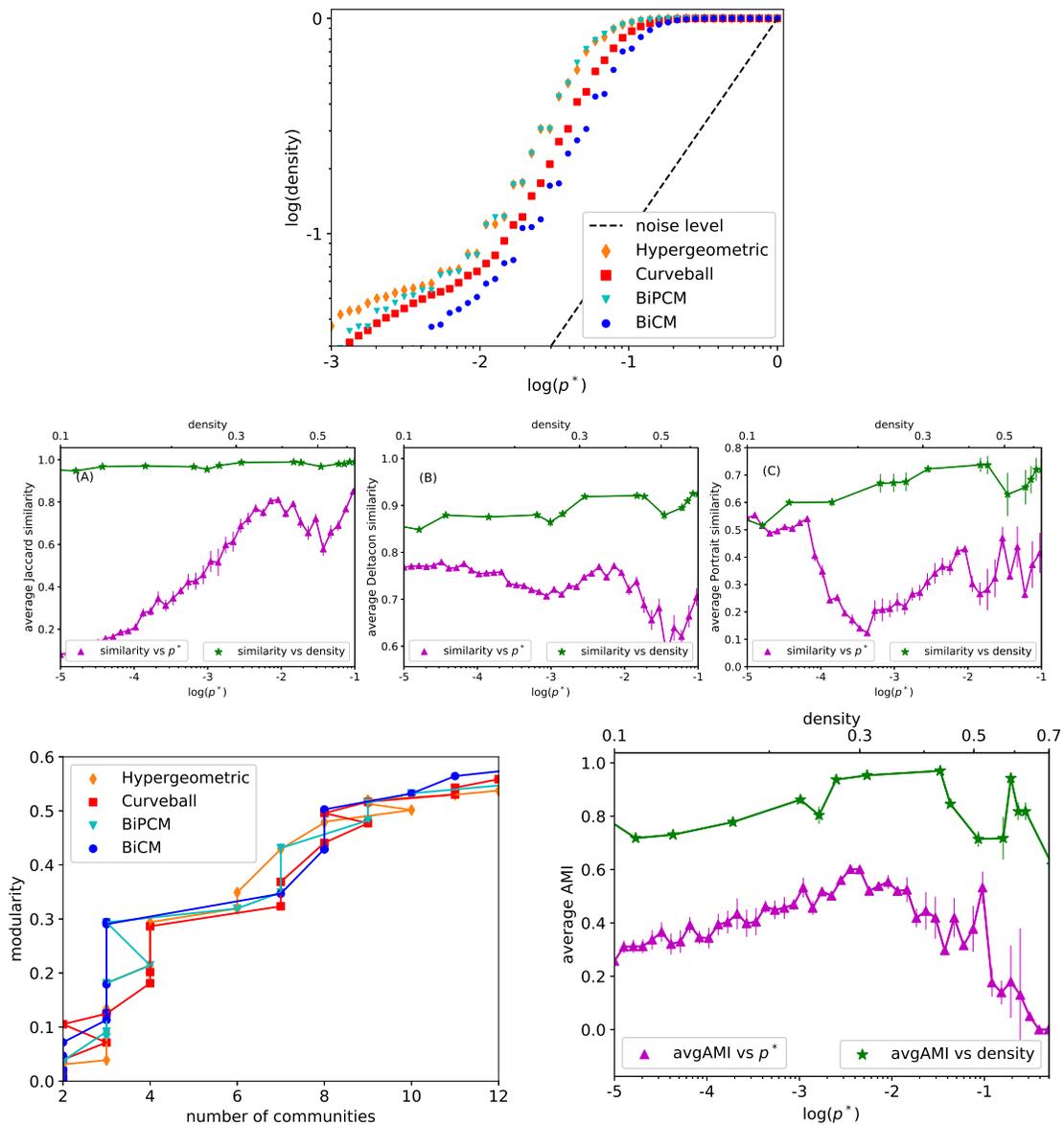

Supplementary Figure 16. *DUTCH* network. Top panel: density $\rho$ of validated links as a function of $p^*$. Middle panels: (A) Jaccard, (B) DeltaCon, (C) Portrait similarity of validated networks, as a function of either $p^*$ or $\rho$. Bottom panels: (left) Modularity vs number of communities; (right) average AMI for partitions of the validated networks as a function of either $p^*$ or $\rho$. Error bars represent standard deviations over choices of null model pairs.

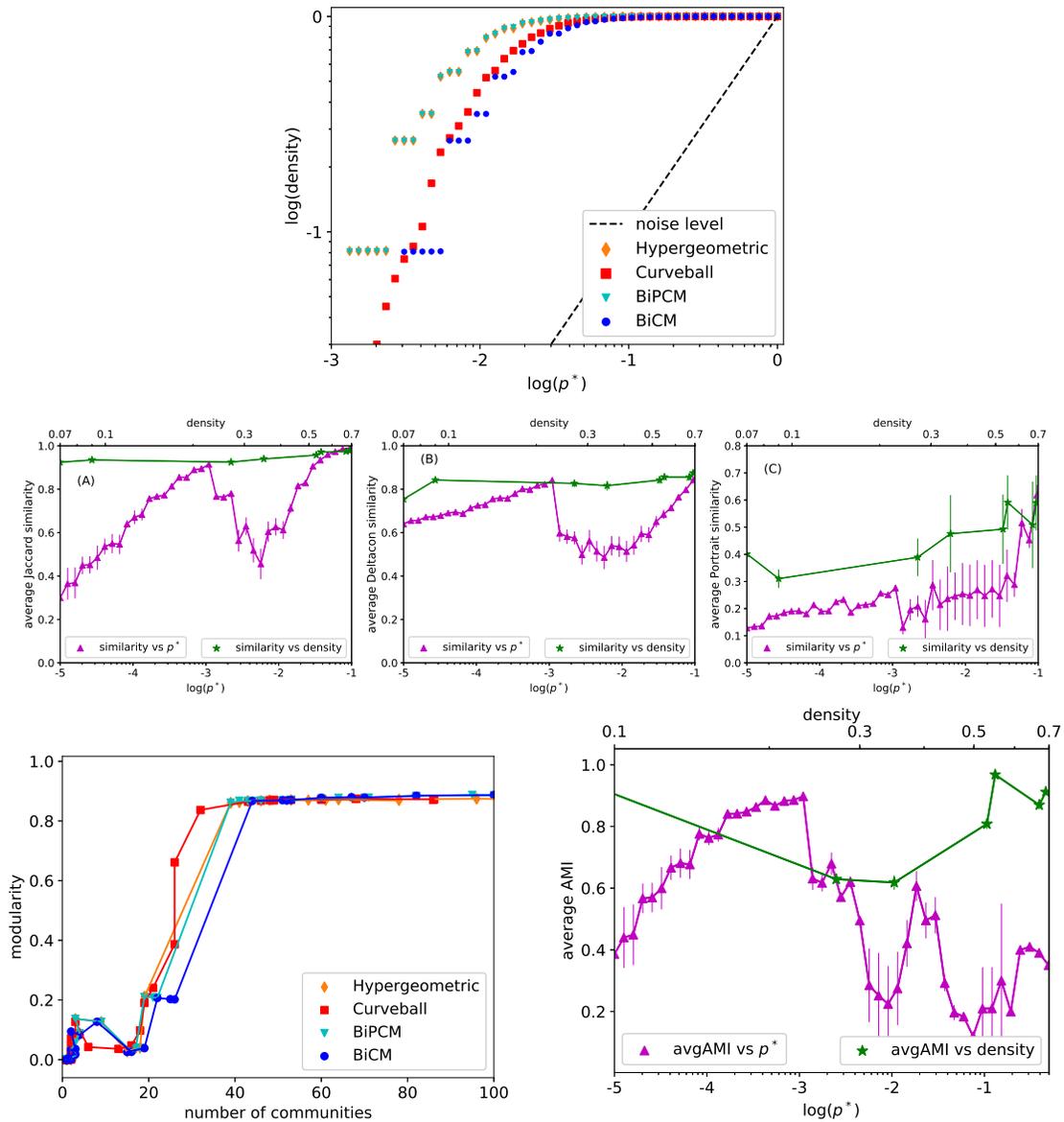

Supplementary Figure 17. *CRIME* network. Top panel: density $\rho$ of validated links as a function of $p^*$. Middle panels: (A) Jaccard, (B) DeltaCon, (C) Portrait similarity of validated networks, as a function of either $p^*$ or $\rho$. Bottom panels: (left) Modularity vs number of communities; (right) average AMI for partitions of the validated networks as a function of either $p^*$ or $\rho$. Error bars represent standard deviations over choices of null model pairs.

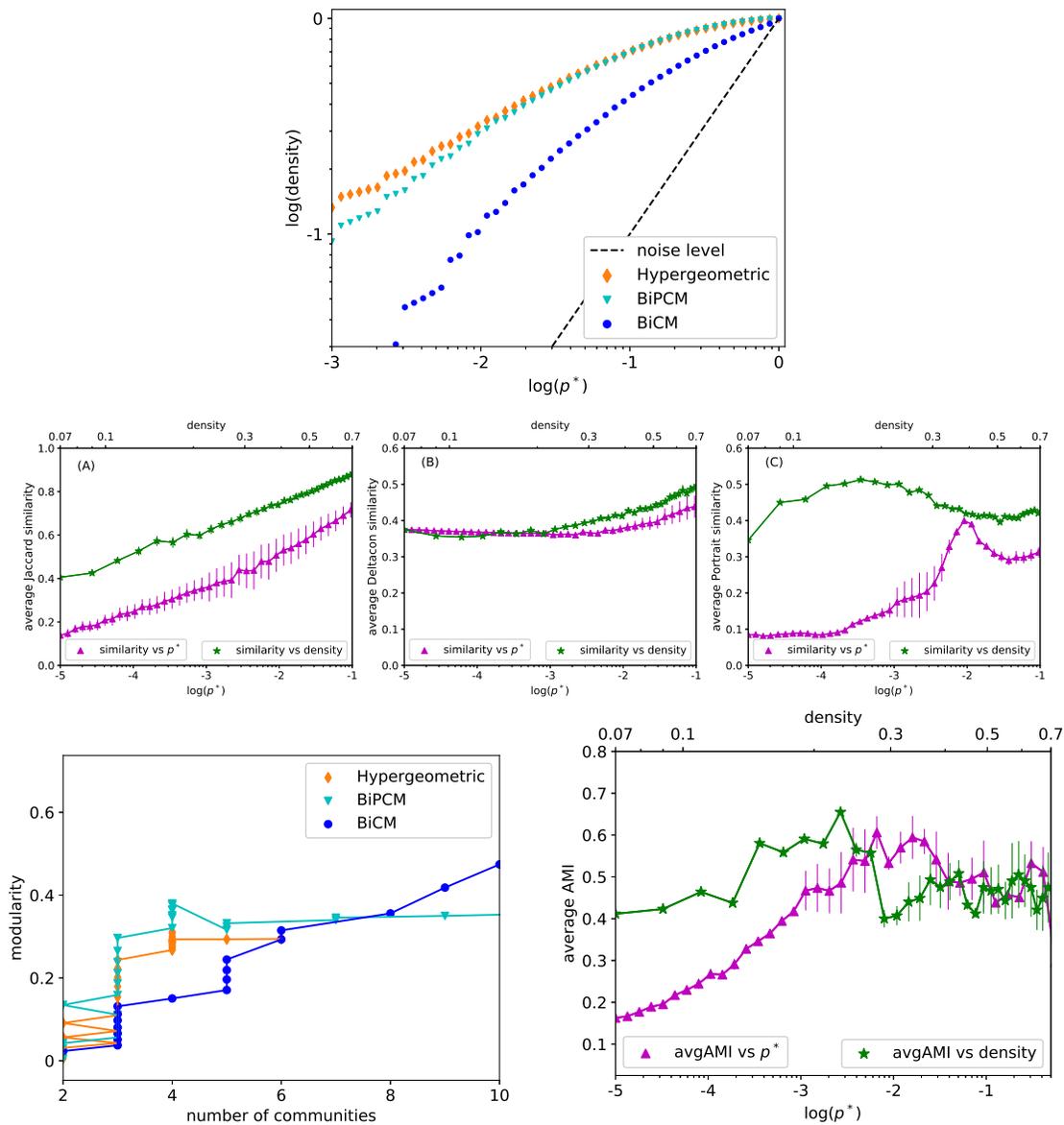

Supplementary Figure 18. *MOVIELENS* network. Top panel: density $\rho$ of validated links as a function of $p^*$. Middle panels: (A) Jaccard, (B) DeltaCon, (C) Portrait similarity of validated networks, as a function of either $p^*$ or $\rho$. Bottom panels: (left) Modularity vs number of communities; (right) average AMI for partitions of the validated networks as a function of either $p^*$ or $\rho$. Error bars represent standard deviations over choices of null model pairs.



\end{document}